\documentclass{article}
\usepackage{amssymb}
\usepackage{amsfonts}
\usepackage{amsmath}

\input{tcilatex}
\begin{document}

\begin{center}
\bigskip

{\large Standard Model Derivation from a 4-d Pseudo-Conformal Field Theory}

{\large \ }\bigskip

C. N. Ragiadakos

IEP, Ministry of Education

email: ragiadak@gmail.com

\bigskip

\textbf{ABSTRACT}
\end{center}

Pseudo-conformal field theory (PCFT) is a 4-d action, which depends on the
lorentzian Cauchy-Riemann (LCR) structure, determined by a tetrad satisfying
precise integrability conditions. This LCR-tetrad defines a class of
Einstein metrics and an electroweak U(2) connection. A static massive and a
massless LCR-manifolds are found. The massive soliton is compatible with the
Kerr-Newman manifold. Its two conjugate LCR-structures have g=2 gyromagnetic
ratio and opposite charges, suggesting their identification with the
electron and positron particles with a naked ring essential singularity.
Their background CP(3) formulation bypasses the Hawking-Penrose singularity
theorems. The massless LCR-manifold does not have a charge, suggesting its
identification with the neutrino. The LCR-structure formalism provides the
particles separated into left and right handed chiral parts, the left and
right columns of the homogeneous coordinates of the grassmannian G(4,2). The
electron LCR-tetrad explicitly provides its gravitational and electroweak
potentials (dressings). Their distributional nature permit us to use the
Bogoliubov causal perturbative approach (improved by Epstein-Glaser and
Scharf et. al. techniques) as a pure mathematical harmonic expansion in the
Gelfand rigged Hilbert-Fock space of tempered distributions of the Poincare
representations (corresponding free fields). This S-matrix computational
procedure in the proper Gelfand triplet, provides the standard model
lagrangian for the electromagnetic, weak and Higgs interactions. The
interacting terms and the relation between the masses and the coupling
constants are implied by the Scharf et. al. operational algorithm on the
free fields. In PCFT the computed gluon potential (static quark dressing)
cannot be treated with the Bogoliubov procedure. Possible solutions of the
dark matter and neutrino mixing problems are discussed.

\bigskip 

\newpage 

\bigskip 

{\LARGE Contents}

\textbf{1. INTRODUCTION}

\textbf{2. GENERAL\ RELATIVITY\ AND\ GRAVITON}

\qquad 2.1 Solving the electron naked singularity problem

\textbf{3. THE VACUUM}

\textbf{4. BOGOLIUBOV'S PERTURBATIVE\ QFT}

\qquad 4.1 Derivation of quantum electrodynamics

\textbf{5. ELECTROWEAK\ GAUGE FIELD}

\qquad 5.1 On the origin of the particle generations

\textbf{6. HADRONIC SECTOR\ AND\ CONFINEMENT}

\textbf{7. BEYOND\ THE\ STANDARD\ MODEL}

\qquad 7.1 On the dark matter and energy

\qquad 7.2 Harmonic expansion in the bounded domain

\qquad \newpage

\bigskip

\renewcommand{\theequation}{\arabic{section}.\arabic{equation}}

\section{INTRODUCTION}

\setcounter{equation}{0}

It is well known that 4-dimensional generally covariant lagrangian models,
based on riemannian geometry, are not renormalizable. Even if they are
endowed with the Weyl symmetry, they turn out not to be compatible with
quantum mechanics, because of the emergence of the product of two Weyl
tensors. It is well understood that lagrangians with second order
derivatives generate negative norm states in the Hilbert space. Hence we
have to look for metric independent lagrangians, which are not topological.

The original idea\cite{RAG1988}$^{,}$\cite{RAG2008b} to study
(Cauchy-Riemann) CR-structure dependent field theories emerged from the
observation that the Polyakov string action 
\begin{equation}
\begin{array}{l}
I_{S}=\frac{1}{2}\int d^{2}\!\xi \ \sqrt{-\gamma }\ \gamma ^{\alpha \beta }\
\partial _{\alpha }X^{\mu }\partial _{\beta }X^{\nu }\eta _{\mu \nu } \\ 
\end{array}
\label{i1}
\end{equation}%
does not essentially depend on the metric $\gamma ^{\alpha \beta }$ of the
2-dimensional surface, because in the light-cone coordinates $(\xi _{-},\
\xi _{+})$ it takes the metric independent form 
\begin{equation}
\begin{array}{l}
I_{S}=\int d^{2}\!z\ \partial _{-}X^{\mu }\partial _{+}X^{\nu }\eta _{\mu
\nu } \\ 
\end{array}
\label{i2}
\end{equation}%
which is not a topological lagrangian. This metric independence is based on
the fundamental property of the 2-dimensional riemannian manifolds\ to admit
a coordinate system $(\xi _{-},\ \xi _{+})$ such that $ds^{2}=2\gamma d\xi
_{+}d\xi _{-}$. This metric independence of the action, without being
topological, is the crucial property of the Polyakov action, which should be
transferred to four dimensions and not the simple Weyl invariance. That is,
the four dimensional analogous symmetry has to be a form of pseudo-conformal
symmetry (Cauchy-Riemann structure) and not the conventional Weyl symmetry.

Four dimensional spacetimes metrics cannot generally take the form (\ref{i2}%
). Only metrics which admit two geodetic and shear free null congruences $%
\ell ^{\mu }\partial _{\mu },\ n^{\mu }\partial _{\mu }$ can take\cite%
{FLAHE1974}$^{,}$\cite{FLAHE1976} the analogous form 
\begin{equation}
\begin{array}{l}
ds^{2}=2g_{a\widetilde{\beta }}dz^{\alpha }dz^{\widetilde{\beta }}\quad
,\quad \alpha ,\widetilde{\beta }=0,1 \\ 
\end{array}
\label{i3}
\end{equation}%
where $z^{b}=(z^{\alpha }(x),z^{\widetilde{\beta }}(x))$ are generally
complex coordinates. In this case we can write down the following metric
independent Yang-Mills-like action 
\begin{equation}
\begin{array}{l}
I_{G}=\int d^{4}\!z\ \sqrt{-g}g^{\alpha \widetilde{\alpha }}g^{\beta 
\widetilde{\beta }}F_{\!j\alpha \beta }F_{\!j\widetilde{\alpha }\widetilde{%
\beta }}+c.c.=\int d^{4}\!z\ F_{\!j01}F_{\!j\widetilde{0}\widetilde{1}}+c.\
c. \\ 
\\ 
F_{j_{ab}}=\partial _{a}A_{jb}-\partial _{a}A_{jb}-\gamma
_{h}\,f_{jik}A_{ia}A_{kb}%
\end{array}
\label{i4}
\end{equation}%
which depends on the CR-structure coordinates, but it does not depend on the
metric.

Notice the similarity of this 4-dimensional action with the 2-dimensional
Polyakov action (\ref{i2}). In the place of the "field" $X^{\mu }$, which is
interpreted as the background 26-dimensional Minkowski spacetime in string
theory, we now have a gauge field $A_{j\nu }$, which we have to interpret as
the gluon, because the field equations generate a linear potential instead
of the Coulomb-like ($\frac{1}{r}$) potential of ordinary Yang-Mills action.

The present action is based on the lorentzian CR-structure\cite{RAG2013b},
which is determined by two real and one complex independent 1-forms ($\ell
,n,m,\overline{m}$), which satisfy the relations 
\begin{equation}
\begin{array}{l}
d\ell =Z_{1}\wedge \ell +i\Phi _{1}m\wedge \overline{m} \\ 
\\ 
dn=Z_{2}\wedge n+i\Phi _{2}m\wedge \overline{m} \\ 
\\ 
dm=Z_{3}\wedge m+\Phi _{3}\ell \wedge n \\ 
\end{array}
\label{i5}
\end{equation}%
where the vector fields $Z_{1\mu }\ ,\ Z_{2\mu }\ $ are real, the vector
field$\ Z_{3\mu }$ is complex, the scalar fields $\Phi _{1}\ ,\ \Phi _{2}$
are real and the scalar field$\ \Phi _{3}$ is complex. This structure
essentially replaces the riemannian structure of the spacetime in the
Einstein general relativity. The form (\ref{i5}) is completely integrable
via the (holomorphic) Frobenious theorem, which implies that the lorentzian
CR-manifold (LCR-manifold) is defined\cite{BAOU} as a 4-dimensional totally
real submanifold of $%
\mathbb{C}
^{4}$ determined by four special (real) functions, 
\begin{equation}
\begin{array}{l}
\rho _{11}(\overline{z^{\alpha }},z^{\alpha })=0\quad ,\quad \rho
_{12}\left( \overline{z^{\alpha }},z^{\widetilde{\alpha }}\right) =0\quad
,\quad \rho _{22}\left( \overline{z^{\widetilde{\alpha }}},z^{\widetilde{%
\alpha }}\right) =0 \\ 
\end{array}
\label{i6}
\end{equation}%
where $\rho _{11}\ ,\ \rho _{22}$ are real and $\rho _{12}$ is a complex
function and $z^{b}=(z^{\alpha },z^{\widetilde{\alpha }}),\ \alpha =0,1$ are
the local structure coordinates in $%
\mathbb{C}
^{4}$. Notice the special dependence of the defining functions on the
structure coordinates. They are not general functions of $z^{b}$. The
separation of chiralities in the standard model is caused to this property.
The LCR-structure is more general than the riemannian structure of general
relativity and permits the invariance of the set of solutions to the
pseudo-conformal transformations (in the E. Cartan and Tanaka terminology)%
\cite{CARTAN}.

The action (\ref{i4}) takes the following generally covariant form 
\begin{equation}
\begin{array}{l}
I_{G}=\int d^{4}\!x\ \sqrt{-g}\ \left\{ \left( \ell ^{\mu }m^{\rho
}F_{\!j\mu \rho }\right) \left( n^{\nu }\overline{m}^{\sigma }F_{\!j\nu
\sigma }\right) +\left( \ell ^{\mu }\overline{m}^{\rho }F_{\!j\mu \rho
}\right) \left( n^{\nu }m^{\sigma }F_{\!j\nu \sigma }\right) \right\} \\ 
\\ 
F_{j\mu \nu }=\partial _{\mu }A_{j\nu }-\partial _{\nu }A_{j\mu }-\gamma
\,f_{jik}A_{i\mu }A_{k\nu }%
\end{array}
\label{i7}
\end{equation}%
where we have to consider the additional action term with the integrability
conditions on the tetrad 
\begin{equation}
\begin{array}{l}
I_{C}=-\int d^{4}\!x\ \{\phi _{0}(\ell ^{\mu }m^{\nu }-\ell ^{\nu }m^{\mu
})(\partial _{\mu }\ell _{\nu })+ \\ 
\\ 
\qquad +\phi _{1}(\ell ^{\mu }m^{\nu }-\ell ^{\nu }m^{\mu })(\partial _{\mu
}m_{\nu })+\phi _{\widetilde{0}}(n^{\mu }\overline{m}^{\nu }-n^{\nu }%
\overline{m}^{\mu })(\partial _{\mu }n_{\nu })+ \\ 
\\ 
\qquad +\phi _{\widetilde{1}}(n^{\mu }\overline{m}^{\nu }-n^{\nu }\overline{m%
}^{\mu })(\partial _{\mu }\overline{m}_{\nu })+c.conj.\}%
\end{array}
\label{i8}
\end{equation}%
These Lagrange multipliers introduce the integrability conditions of the
tetrad and make the complete action $I=I_{G}+I_{C}$ self-consistent and the
usual quantization techniques may be used\cite{RAG1992}. The action is
formally renormalizable\cite{RAG2008a}, because it is dimensionless and
metric independent. Its path-integral quantization is also formulated\cite%
{RAG2017} as functional summation of open and closed 4-dimensional
lorentzian CR-manifolds in complete analogy to the summation of
2-dimensional surfaces in string theory\cite{POL}. These transition
amplitudes of a quantum theory of LCR-manifolds provides the self-consistent
algorithms for the computation of the physical quantities. But
unfortunately, I have not yet found a method to compute these functional
integrals, therefore I will use a "solitonic" technique\cite{RAG1991}$^{,}$%
\cite{RAG1999}, which appears in the linearized Einstein gravity
approximation. The present paper should be considered as a continuation of
the last one\cite{RAG2017}, which will be called [paper I]. In this [paper
I] the reader may find a review of the (lorentzian) LCR-structure\cite%
{RAG2013b}, which is the fundamental mathematical structure of the present
pseudo-conformal field theory (PCFT). The properties of this structure will
be used in the present work without proof, in order to facilitate the
understanding of the general framework of the procedure.

The LCR-structure defining relations are invariant under the following
tetrad-Weyl transformations

\begin{equation}
\begin{tabular}{l}
$\ell _{\mu }^{\prime }=\Lambda \ell _{\mu }\quad ,\quad n_{\mu }^{\prime
}=Nn_{\mu }\quad ,\quad m_{\mu }^{\prime }=Mm_{\mu }$ \\ 
\\ 
$\ell ^{\prime \mu }=\frac{1}{N}\ell ^{\mu }\quad ,\quad n^{\prime \mu }=%
\frac{1}{\Lambda }n^{\mu }\quad ,\quad m^{\prime \mu }=\frac{1}{\overline{M}}%
m^{\mu }$ \\ 
\end{tabular}
\label{i9}
\end{equation}%
with non-vanishing $\Lambda \ ,\ N\ ,\ M$. I point out that we have not yet
introduced a metric. The tetrad with upper and lower indices is simply a
basis of tangent and cotangent spaces. But the tetrad does define a class $%
[g_{\mu \nu }]$ of symmetric tensors \ 
\begin{equation}
\begin{array}{l}
g_{\mu \nu }=\ell _{\mu }n_{\nu }+\ell _{\nu }n_{\mu }-m_{\mu }\overline{m}%
_{\nu }-m_{\nu }\overline{m}_{\mu } \\ 
\end{array}
\label{i10}
\end{equation}%
Every such tensor may be used as a metric to build up the riemannian
geometry of general relativity, because its local signature is ($1,-1,-1,-1$%
). But this form always admits two geodetic and shear free null congruences
and hence it does not cover all the metrics of general relativity. I think
that this restriction will not cause any phenomenological problem to the
model, because all the known gravitational objects do admit such
congruences. Besides, notice that the tetrad-Weyl symmetry (\ref{i9}) is
larger than the well known metric-Weyl symmetry of the quadratic Weyl tensor
lagrangian. Because of these symmetries the PCFT is renormalizable\cite%
{RAG2008a}.

The conventional solitons\cite{FELS1981} are defined as classical solutions
with finite mass determined via the energy-momentum conserved current, which
besides, are "protected" to deform to the vacuum configuration by
topological invariants. In the present context the soliton is a LCR-manifold
which in the linearized Einstein ($g_{\mu \nu }$) gravity approximation has
finite mass computed from the conserved gravitational source. The
LCR-structure is "protected" by topological invariants and/or its relative
invariants defined\cite{RAG2013b}$^{,}$\cite{RAG2013a} from the
non-vanishing $\Phi _{i}$.

The standard model gauge field will be explicitly defined from the
LCR-tetrad. The electron solitonic LCR-surface and its electromagnetic, weak
interactions and Higgs potentials will be also determined. The construction
of the effective "quantum field model" will be performed using the
Bogoliubov-Medvedev-Polivanov (BMP)\cite{BOG1975}$^{,}$\cite{BOG1980}
axiomatic framework. The S-matrix is function of the free effective fields
and it is constructed order by order. Starting from the classical
interaction (correspondence principle) the effective QFT will be built up
introducing the necessary additional terms and conditions such that the
final action to be well defined. The incorporation of the Epstein-Glaser (%
\cite{EpGl}) regularization procedure and the Scharf $Q$-charge technique%
\cite{Sch1}$^{,}$\cite{Sch2} provides the standard model action.

In order to make things as simple as possible I will proceed step by step.
In section II, the linearized Einstein gravity approximation is described
and the graviton and its source current is defined. In the subsection we
prove how the LCR-structure solves the puzzle of the naked singularity of
Kerr-Newman manifold with the electron mass, charge and spin parameters. It
makes clear that LCR-structure may bypass the Hawking-Penrose singularity
theorem in riemannian geometry, which did not permit to relate the electron
with the Kerr-Newman manifold. In section III, the degenerate LCR-structure
will be studied, which is assumed to be the vacuum\cite{RAG2017} of the
effective QFT. I repeat this analysis, already done in [paper I], in order
to make clear not only the vacuum conservation of the Poincar\'{e} group,
but also the deep LCR-structure origin of the chirality, which is
fundamental in the standard model. The left and right separation of the
infinite group of pseudo-conformal transformations (in the E. Cartan and
Tanaka terminology) is a fundamental property of the LCR-structure. We will
show that outside of the distributional singularity of the solitonic
potentials (dressings) the symmetry group is the Poincar\'{e} group which is
essential to consider the Bogoliubov perturbative expansion as a Poincar\'{e}
harmonic expansion. In section IV, the static LCR-manifold is explicitly
derived, which is identified with the electron. Its complex conjugate is
identified with the positron, because they are found to have opposite
charges. The electromagnetic potentials are defined by a self-dual 2-form,
which happens to be locally integrable. The photon interacts with the
electron and positron currents with opposite charges. The corresponding
metric is the Kerr-Newman manifold, where the electromagnetic potential is
found to be proportional to the geodetic and shear-free null vector $\ell
_{\mu }$. It is not a computational accident, because the defined (in
section V) electroweak connection extends this relation to all the
electroweak $U(2)$ gauge fields. A subsection is devoted to describe the BMP
procedure, which suggests that the order-by-order implied formal
electromagnetic potential of the electron should be identified to the
corresponding expansion of $A_{\mu }$ relative to the Kerr-Newman parameter $%
a$. The left and right chiralities emerge from the homogeneous coordinates
of grassmannian manifold $G_{4,2}$ of the lines of $CP^{3}$. In order to
have a physical intuition through all the\ mathematical steps, I will use
the generally complex Newman trajectories\cite{NEWM1973}$^{,}$\cite{NEWM2004}
to determine the LCR-structure. They are essentially characteristic
properties of the ruled surfaces\cite{RAG2021} of $CP^{3}$.

In section V, the massless stationary LCR-structure, determined from a
reducible quadratic Kerr polynomial, is computed. It has a clear asymmetry
between the left and right handed parts of the $G_{4,2}$ homogeneous
coordinates. This LCR-structure is identified with the neutrino, because
only its left-handed part is not degenerate (trivial), while the right part
is that of the trivial vacuum. Besides, its integrable self-dual 2-form does
not define any charge. The general LCR-tetrad ($\ell ,m;n,\overline{m}$) can
be adapted into the [2.78 of paper I] $U(2)$ group and the electroweak
potentials are computed in the electron static LCR-structure. I think that
this discovery fulfills the Einstein's goal towards a geometric unified
theory.

In section VI, the gauge field compatible with the symmetries of
LCR-structure is identified with the gluon field and a confining gauge
potential is computed in the static "quark" LCR-structure. But the gluon
propagator does not coincide with that of conventional quantum
chromodynamics, because it does not permit an expansion in the spin
parameter $a$. In section VII the dark matter and the neutrino mixing
problems proposing possible solutions.

The general result is that the gravitational, electromagnetic, weak and
Higgs interactions are exactly derived, but the gluon propagator of QCD must
be replaced with the present confining gauge field of PCFT. The interested
reader may find more details in my frequently updated Research eBook\cite%
{RAG2021}.

\section{GENERAL\ RELATIVITY\ AND\ GRAVITON}

\setcounter{equation}{0}

The geometric dynamical variables of the present model are the two real and
the one complex vectors ($\ell ,n,m,\overline{m}$), which define the
lorentzian CR-structure. They determine the symmetric tensor (\ref{i10}),
which will be identified with the Einstein metric. But these metrics are not
invariant under the tetrad-Weyl symmetry (\ref{i9}) of the LCR-structure.
Therefore a LCR-structure defines a class of metrics [$g_{\mu \nu }$]. Two
metrics related by a tetrad-Weyl symmetry belong to the same class.

On the other hand the local $SO(1,3)$ symmetry\cite{CHAND} of this symmetric
tensor does not preserve the geodetic and shear free conditions ($\kappa
=\sigma =\lambda =\nu =0$)\cite{FLAHE1976} of two null vectors, which are
equivalent to the definition (\ref{i5}) of the LCR-structure. Therefore it
is not a symmetry of the present fundamental (geometric) LCR-structure.

The tetrad ($\ell ,n,m,\overline{m}$) may be used to write down other
symmetric tensors, but it is the form (\ref{i10}) that makes the lagrangian (%
\ref{i4}) of the model metric independent. Besides,, this precise metric
form permits us to define the flat spacetimes and asymptotically flat
spacetimes at null infinity using directly the LCR-structure solutions.

If [$g_{\mu \nu }$] contains the Minkowski metric, the LCR-structure is
determined\cite{RAG2013b}$^{,}$\cite{RAG2013a} by an element of the $G_{4,2}$
grassmannian projective space with homogeneous coordinates $X^{mi}$, which
belong to the Kerr surface $K(X^{i})\ ,i=1,2$ of $CP^{3}$, and such that \ 
\begin{equation}
\begin{array}{l}
X^{\dag }%
\begin{pmatrix}
0 & \mathbf{1} \\ 
\mathbf{1} & 0%
\end{pmatrix}%
X=0 \\ 
\end{array}
\label{g1}
\end{equation}%
Notice that the inverse is also true. These LCR-structures always define a
Minkowski metric on the "real axis" of the Siegel domain, up to the
appearance of singularities. This is the characteristic (Shilov) boundary of
the $SU(2,2)$ symmetric classical domain. This manifold generally admits an
infinite number of LCR-structures locally determined by an irreducible or
reducible Kerr homogeneous polynomials. The characteristic property of these
"flat" LCR-structures is that $X$ admits the form \ 
\begin{equation}
\begin{array}{l}
X=%
\begin{pmatrix}
\lambda \\ 
-ix\lambda%
\end{pmatrix}
\\ 
\end{array}
\label{g2}
\end{equation}%
where $x$ is a hermitian, $\lambda $ is a complex $2\times 2$ matrix, and
the two columns are determined by the Kerr homogeneous polynomials. That is,
their "left" and "right" parts decuple. The LCR-structures, which cannot
take the form (\ref{g1}) will be called "curved", and vice-versa their
approximations restricted to these terms will be called the "flatprints" of
a generic LCR-structure.

In the linearized Einstein gravity approximation\cite{MTW}, we find the
following linearized gravity relations in the limit \ 
\begin{equation}
\begin{array}{l}
g_{\mu \nu }=\eta _{\mu \nu }+kh_{\mu \nu }+O(k^{2}) \\ 
\\ 
\widehat{R}_{\nu \rho \sigma \tau }=\underset{k\rightarrow 0}{\lim
(k^{-1}R_{\nu \rho \sigma \tau }})=2\partial _{\lbrack \nu }\partial
_{|[\sigma }h_{\tau ]|\rho ]} \\ 
\end{array}
\label{g3}
\end{equation}%
for the curvature tensor. The second Bianchi identity takes the form \ 
\begin{equation}
\begin{array}{l}
\partial _{\lbrack \mu }\widehat{R}_{\nu \rho ]\sigma \tau }=0 \\ 
\\ 
\partial _{\mu }\widehat{R}_{\ \nu \rho \sigma }^{\mu }=\partial _{\rho }%
\widehat{R}_{\nu \sigma }-\partial _{\rho }\widehat{R}_{\nu \sigma } \\ 
\end{array}
\label{g4}
\end{equation}%
where the covariant derivative becomes minkowiskian and $[...]$ denotes
antisymmetrization. They imply the conservation condition of the Einstein
tensor \ 
\begin{equation}
\begin{array}{l}
\partial _{\mu }\widehat{E}_{\ \nu }^{\mu }=\partial _{\mu }[\widehat{R}_{\
\nu }^{\mu }-\frac{1}{2}\delta _{\nu }^{\mu }\widehat{R}]=0 \\ 
\end{array}
\label{g5}
\end{equation}%
This means that the Einstein tensor is conserved in the linearized Einstein
gravity limit. Besides, in the empty space it becomes the free wave equation
of a massless spin-2 particle.

Recall that in relativistic quantum field theory the field, which satisfies
a spin-s free wave equation, describes\cite{TUNG} a representation of the
Poincar\'{e} group i.e. a spin-s particle, and vice-versa, a spin-s particle
is described by a field representation of the Poincar\'{e} group, which
satisfies the free wave equation. Hence, in the present model, Einstein's
general relativity naturally emerges. The existence of a graviton is simply
implied in the weak gravity limit.

In the Penrose spinorial formalism\cite{P-R1984}, the linearized Bianchi
identity takes the form \ 
\begin{equation}
\begin{array}{l}
\partial _{B^{\prime }}^{A}\widehat{\Psi }_{ABCD}=\partial _{(B}^{A^{\prime
}}\widehat{\Phi }_{CD)A^{\prime }B^{\prime }} \\ 
\end{array}
\label{g6}
\end{equation}%
where $(...)$ denotes symmetrization. The left-hand side contains the Weyl
tensor and the right-hand side of the relation contains the Ricci tensor,
which describes the sources. It is considered as the graviton wave equation.

I have already pointed out that PCFT defines only metrics which admit
geodetic and shear-free congruences. In this case the flags $\lambda ^{A}$
of the LCR-structure tetrad must satisfy the condition \ 
\begin{equation}
\begin{array}{l}
\Psi _{ABCD}\lambda ^{A}\lambda ^{B}\lambda ^{C}\lambda ^{D}\simeq k\widehat{%
\Psi }_{ABCD}\widehat{\lambda }^{A}\widehat{\lambda }^{B}\widehat{\lambda }%
^{C}\widehat{\lambda }^{D}+O(k^{2})=0 \\ 
\end{array}
\label{g7}
\end{equation}%
where the linearized gravity approximation has been also considered. The
Weyl tensor of the Minkowski spacetime vanishes. Therefore the gravitational
content remains in $\widehat{\Psi }_{ABCD}$ and $\widehat{\lambda }^{A}(x)$
are the flags of the flatprint of the LCR-structure tetrad.

Notice that the gravitational singularities are locally determined by the
zeroes and infinities of the metric and its inverse, which cannot be removed
by a coordinate change. These metric-singularities essentially coincide with
the LCR-structure singularities, because $\det (g_{\mu \nu })=-C(\det [\ell
_{\mu },n_{\mu },m_{\mu },\overline{m_{\mu }}])^{2}$. Hence, the singularity
sources of the gravitational radiation coincide with the singularities of
the LCR-structure, which defines the corresponding metric. But in the
linearized Einstein gravity limit, the singular region of the form (\ref{g2}%
) is determined by the Kerr functions. These are the regions where two roots
of the homogeneous functions $K_{i}(X^{i})\ ,i=1,2$ coincide. More details
will be given in the next sections, where the same method will be used to
define the electromagnetic potential.

Newman has found\cite{NEWM1973} that the Kerr function condition (for a null
congruence to be geodetic and shear-free) may be replaced with a (generally
complex) trajectory $\xi ^{a}(\tau )$. In the present case of the
LCR-structure formalism, this is done by assuming that the $G_{4,2}$ two
homogeneous coordinates $i=1,2$ must have the form \ 
\begin{equation}
\begin{array}{l}
X^{i}=%
\begin{pmatrix}
\lambda ^{i} \\ 
-i\xi _{(i)}(\tau _{i})\lambda ^{i}%
\end{pmatrix}
\\ 
\\ 
\xi _{(i)}(\tau _{i})=\xi _{(i)}^{a}(\tau _{i})\sigma ^{b}\eta _{ab} \\ 
\end{array}
\label{g10}
\end{equation}%
where $\sigma ^{b}$and $\eta _{ab}$ are the Pauli matrices and the Minkowski
metric respectively. Here, I have to point out that the consideration of two
generally different complex Kerr homogeneous functions is somehow
misleading. In conventional algebraic geometry the notion of reducible
polynomial is used. The irreducible Kerr polynomial \ 
\begin{equation}
\begin{array}{l}
K(Z)=Z^{1}Z^{2}-Z^{0}Z^{3}+2aZ^{0}Z^{1} \\ 
\end{array}
\label{g10a}
\end{equation}
of the electron LCR-structure is equivalent with the complex trajectory $\xi
^{a}=(\tau ,0,0,ia)$. The complex trajectory is a characteristic property of
the ruled surfaces\cite{RAG2021} of $CP^{3}$.

The flatprint LCR-structure coordinates are determined by the condition \ 
\begin{equation}
\begin{array}{l}
(x-\xi _{(i)}(\tau _{i}))\lambda ^{i}=0 \\ 
\end{array}
\label{g11}
\end{equation}%
that admits one non-vanishing solution for every column $i=1,2$ of the
homogeneous coordinates of $G_{4,2}$. This is possible if \ 
\begin{equation}
\begin{array}{l}
\det (x-\xi _{(i)}(\tau _{i}))=\eta _{ab}(x^{a}-\xi _{(i)}^{a}(\tau
_{i}))(x^{b}-\xi _{(i)}^{b}(\tau _{i}))=0 \\ 
\end{array}
\label{g12}
\end{equation}%
which gives the two solutions $z^{0}=\tau _{1}(x)$ and $z^{\widetilde{0}%
}=\tau _{2}(x)$. The other structure coordinates are $z^{1}=\frac{\lambda
^{11}}{\lambda ^{01}}$ and $z^{\widetilde{1}}=-\frac{\lambda ^{02}}{\lambda
^{12}}$ where 
\begin{equation}
\begin{array}{l}
\lambda ^{Aj}=%
\begin{pmatrix}
(x^{1}-ix^{2})-(\xi _{(j)}^{1}(\tau _{j})-i\xi _{(j)}^{2}(\tau _{j})) \\ 
(x^{0}-x^{3})-(\xi _{(j)}^{0}(\tau _{j})-\xi _{(j)}^{3}(\tau _{j}))%
\end{pmatrix}
\\ 
\end{array}
\label{g13}
\end{equation}%
Notice that the trajectory technique for computation of the structure
coordinates incorporates the notion of the classical causality, which is
apparently respected by (\ref{g12}).

The singularity of the flatprint LCR-structure occurs at $\det [\lambda
^{A1}(x),\lambda ^{A2}(x)]=0$. Recall that the left and right columns of the
homogeneous coordinates of $G_{4,2}$ may be determined ("move") with
different trajectories, if the corresponding homogeneous Kerr polynomial is
reducible. In the simple case when both move with the same trajectory $\xi
^{a}(\tau )=(\tau ,\xi ^{1}(\tau ),\xi ^{2}(\tau ),\xi ^{3}(\tau ))$, the
singularity occurs at $\tau _{1}(x)=\tau _{2}(x)$, which is 
\begin{equation}
\begin{array}{l}
(x^{i}-\xi ^{i}(t))(x^{j}-\xi ^{j}(t))\delta _{ij}=0 \\ 
\end{array}
\label{g14}
\end{equation}%
If $\xi _{R}^{i}$ and $\xi _{I}^{i}$ are the real and imaginary parts of the
trajectory, we find that the locus of the solitonic LCR-structure is 
\begin{equation}
\begin{array}{l}
(x^{i}-\xi _{R}^{i}(t))(x^{j}-\xi _{R}^{j}(t))\delta _{ij}-\xi
_{I}^{i}(t)\xi _{I}^{j}(t)\delta _{ij}=0 \\ 
\\ 
(x^{i}-\xi _{R}^{i}(t)\xi _{I}^{j}(t)\delta _{ij}=0 \\ 
\end{array}
\label{g15}
\end{equation}%
Note that if $\xi _{I}^{j}(t)$ is bounded, the LCR-structure may be
interpreted as a soliton with trajectory $\xi _{R}^{i}(t)$ and a locus at
the perimeter of the circle of radius $(\xi _{I}^{i}(t))^{2}$ around its
trajectory. This locus (a two dimensional surface) is a singularity of the
gravitational potential and a source of the corresponding gravitational
radiation, but it is not a singularity of the LCR-structure viewed as a
surface of the $G_{4,2}$ grassmannian, because the matrix $X^{mi}$ has not
rank two at this surface.

\subsection{Solving the electron naked singularity "problem"}

We saw that the LCR-structure implies Einstein's general relativity based on
the metric geometric structure. But there is an essential difference. The
LCR-structure bypasses the naked singularity problem of the Kerr-Newman
metric. This metric admits two geodetic and shear free null congruences,
which are related with the Kerr polynomial (\ref{g10a}). It also admits two
commuting killing vectors, which are identified with the time-translation
and $z$-rotation generators of the Poincar\'{e} group. Carter's\cite%
{CART1968} discovery that the gyromagnetic ratio of the Kerr-Newman manifold
is fermionic (that of the electron $g=2$)\cite{N-W1974} shocked the
community of general relativists. Many tried to identify the Kerr-Newman
spacetime with the electron without success, because the electron constants
imply the existence of a naked singularity in the Kerr-Newman spacetime.

The electron mass $M_{e}$, charge $e^{2}$ and spin parameter $a$ have the
values%
\begin{equation}
\begin{array}{l}
M=\frac{M_{e}}{M_{P}}=4.18\ast 10^{-23} \\ 
e^{2}=\frac{q^{2}}{4\pi \varepsilon _{0}\hbar c}=\frac{1}{137} \\ 
a=\frac{\hbar }{2M_{e}}=2.09\ast 10^{23} \\ 
\\ 
a^{2}>>e^{2}>>M^{2}%
\end{array}
\label{g16}
\end{equation}%
Hence $a^{2}+e^{2}-M^{2}>0$, and the electron metric has an essential naked
singularity, which is not permitted in riemannian geometry. This is a
problem for general relativity, because its fundamental quantity, the
metric, does not "see" the algebraic structure. It is known (and well
described in many books of general relativity) that its analytic extension
has two sheets $x^{b}$ and $x^{\prime b}$ which are determined by the two
roots%
\begin{equation}
\begin{array}{l}
r=\pm \left\{ \frac{(x^{1})^{2}+(x^{2})^{2}+(x^{3})^{2}-a^{2}}{2}+\sqrt{[%
\frac{(x^{1})^{2}+(x^{2})^{2}+(x^{3})^{2}-a^{2}}{2}]^{2}+a^{2}(x^{3})^{2}}%
\right\} ^{\frac{1}{2}} \\ 
\end{array}
\label{g17}
\end{equation}%
These two surfaces constitute the boundary $U(2)$ of the bounded realization
of the $SU(2,2)$ classical domain and their correspondence is the well known
Cayley transformation. The spinorial electron naked singularity in $U(2)$
universe can be properly incorporated in PCFT, while it is rejected as
"unphysical" by the riemannian formalism. In the context of the unbounded
realization this may be studied using the following LCR-ray tracing.

We have already found that in the flatprint electron LCR-structure the
structure coordinates are%
\begin{equation}
\begin{array}{l}
z^{0}=t-r+ia\cos \theta \quad ,\quad z^{1}=e^{i\varphi }\tan \frac{\theta }{2%
} \\ 
z^{\widetilde{0}}=t+r-ia\cos \theta \quad ,\quad z^{\widetilde{1}}=\frac{r+ia%
}{r-ia}e^{-i\varphi }\tan \frac{\theta }{2} \\ 
\end{array}
\label{g18}
\end{equation}%
and the cartesian coordinates are%
\begin{equation}
\begin{array}{l}
x^{0}=t \\ 
x^{1}+ix^{2}=(r-ia)\sin \theta e^{i\varphi } \\ 
x^{3}=r\cos \theta \\ 
\\ 
r^{4}-[(x^{1})^{2}+(x^{2})^{2}+(x^{3})^{2}-a^{2}]r^{2}-a^{2}(x^{3})^{2}=0%
\end{array}
\label{g19}
\end{equation}%
Then $L^{\mu }\partial _{\mu }z^{\alpha }=0$ implies that the outgoing $%
L^{\mu }$ integral curves (rays) are determined by the surfaces%
\begin{equation}
\begin{array}{l}
s_{1}:=t-r\quad ,\quad s_{2}:=\theta \quad ,\quad s_{3}:=\varphi \\ 
\end{array}
\label{g20}
\end{equation}

Assuming the caustic coordinates ($r,s_{1},s_{2},s_{3}$), which have the
property ($0,s_{1},\frac{\pi }{2},s_{3}$) to be on the caustic. In this
caustic coordinate system the LCR-rays are traced by the relation%
\begin{equation}
\begin{array}{l}
x_{L}^{0}(r)=s_{1}+r \\ 
x_{L}^{1}(r)=(r\cos \varphi +a\sin \varphi )\sin \theta \\ 
x_{L}^{2}(r)=(r\sin \varphi -a\cos \varphi )\sin \theta \\ 
x_{L}^{3}(r)=r\cos \theta \\ 
\\ 
Jacobian=[r^{2}+a^{2}\cos ^{2}\theta ]\sin \theta%
\end{array}
\label{g21}
\end{equation}%
The source of the LCR-rays are at $r=0$, i.e.%
\begin{equation}
\begin{array}{l}
x_{L}^{0}(0)=s_{1} \\ 
x_{L}^{1}(0)=a\sin \varphi \sin \theta \\ 
x_{L}^{2}(0)=-a\cos \varphi \sin \theta \\ 
x_{L}^{3}(0)=0%
\end{array}
\label{g22}
\end{equation}%
the disk found above. Notice that the rays with $s_{2}:=\theta \neq \frac{%
\pi }{2}$ pass through the disk.

The $N^{\mu }\partial _{\mu }z^{\widetilde{\alpha }}=0$ implies that its
incoming $N^{\mu }$ rays are determined by the surfaces%
\begin{equation}
\begin{array}{l}
s_{1}^{\prime }:=t+r\quad ,\quad s_{2}^{\prime }:=\theta \quad ,\quad
s_{3}^{\prime }:=\varphi +\arctan \frac{2ar}{a^{2}-r^{2}} \\ 
\end{array}
\label{g23}
\end{equation}%
Then we find the congruence%
\begin{equation}
\begin{array}{l}
x_{N}^{0}(r)=s_{1}^{\prime }-r \\ 
x_{N}^{1}(r)=[r\cos s_{3}^{\prime }-a\sin s_{3}^{\prime }]\sin \theta \\ 
x_{N}^{2}(r)=[r\sin s_{3}^{\prime }+a\cos s_{3}^{\prime }]\sin \theta \\ 
x_{N}^{3}(r)=r\cos \theta \\ 
\\ 
Jacobian=[r^{2}+a^{2}\cos ^{2}\theta ]\sin \theta%
\end{array}
\label{g24}
\end{equation}%
As expected the velocities $\overset{.}{x}_{L}^{i}(t)$ and $\overset{.}{x}%
_{N}^{i}(t)$ have asymptotically opposite radial directions.

We will now show that the origin of the essential singularity of the Kerr
manifold is the intersection of the two sheets of the static electron
regular quadric (in the unbounded Siegel realization) (\ref{g10a}) of $%
CP^{3} $. In the flatprint case we have 
\begin{equation}
\begin{array}{l}
X^{0}=1\quad ,\quad X^{1}=\lambda \quad ,\quad
X^{2}=-i[(x^{0}-x^{3})-(x^{1}-ix^{2})\lambda ] \\ 
X^{3}=-i[-(x^{1}+ix^{2})+(x^{0}+x^{3})\lambda ] \\ 
\end{array}
\label{g25}
\end{equation}%
and the Kerr polynomial and its two solutions are 
\begin{equation}
\begin{array}{l}
(x^{1}-ix^{2})\lambda ^{2}+2(x^{3}-ia)\lambda -(x^{1}+ix^{2})=0 \\ 
\lambda _{1,2}=\frac{-(x^{3}-ia)\pm \sqrt{\Delta }}{x-iy}\quad ,\quad \Delta
=(x^{1})^{2}+(x^{2})^{2}+(x^{3})^{2}-a^{2}-2iax^{3} \\ 
\end{array}
\label{g26}
\end{equation}%
where $\lambda _{1,2}$\ are the two values of $\lambda $ on the two sheets
of the quadric. The intersection curve of these two sheets is 
\begin{equation}
\begin{array}{l}
\Delta =(x^{1})^{2}+(x^{2})^{2}+(x^{3})^{2}-a^{2}-2iax^{3}=0 \\ 
\\ 
x^{3}=0\quad ,\quad (x^{1})^{2}+(x^{2})^{2}=a^{2}%
\end{array}
\label{g27}
\end{equation}%
which, after the LCR projection to $%
\mathbb{R}
^{4}$, becomes the singularity ring of the electron (Kerr-Newman) manifold.
Notice that the quadratic surface is regular and the intersection of the two
branches is implied by the projection. The points of the algebraic
intersection curve (the branch curve) of the (regular) quadric of $CP^{3}$
are regular points like any other point of the quadric.

In [paper I] we saw that the bounded realization of a flat LCR-manifold is $%
U(2)$, which is covered by two $%
\mathbb{R}
^{4}$ sheets through the Cayley $2\rightarrow 1$ transformation%
\begin{equation}
\begin{array}{l}
For\ s:=R_{0}\frac{\sin \rho }{\cos \tau +\cos \rho }>0 \\ 
x^{0}=T_{0}\frac{\sin \tau }{\cos \tau +\cos \rho } \\ 
x^{1}+ix^{2}=R_{0}\frac{\sin \rho }{\cos \tau +\cos \rho }\sin \sigma \
e^{i\chi } \\ 
x^{3}=R_{0}\frac{\sin \rho }{\cos \tau +\cos \rho }\cos \sigma%
\end{array}
\label{g28}
\end{equation}%
and the second $%
\mathbb{R}
^{4}$ is identified with $s<0$,%
\begin{equation}
\begin{array}{l}
For\ s:=R_{0}\frac{\sin \rho }{\cos \tau +\cos \rho }<0 \\ 
x^{\prime 0}=T_{0}\frac{\sin \tau }{\cos \tau +\cos \rho } \\ 
x^{\prime 1}+ix^{\prime 2}=-R_{0}\frac{\sin \rho }{\cos \tau +\cos \rho }%
\sin \sigma \ e^{i\chi } \\ 
x^{\prime 3}=-R_{0}\frac{\sin \rho }{\cos \tau +\cos \rho }\cos \sigma%
\end{array}
\label{g29}
\end{equation}%
The constants $T_{0}$ and $R_{0}$\ are related to the time and space sizes.
Notice that this is the Penrose artificial compactification of the Minkowski
spacetime, but in the context of PCFT, this is implied by the formalism
itself. In the case of the Penrose artificial compactification these two
sheets $s\gtrless 0$ communicate through the scri+ and scri- infinities. In
the case of the electron flatprint LCR-structure, these two sheets
communicate through the glued two discs $(x^{1})^{2}+(x^{2})^{2}<a^{2}$ too,
because we may assume%
\begin{equation}
\begin{array}{l}
r=+\left\{ \frac{s^{2}-a^{2}}{2}+\sqrt{[\frac{s^{2}-a^{2}}{2}%
]^{2}+a^{2}(x^{3})^{2}}\right\} ^{\frac{1}{2}}\ for\ s>0 \\ 
r=-\left\{ \frac{s^{2}-a^{2}}{2}+\sqrt{[\frac{s^{2}-a^{2}}{2}%
]^{2}+a^{2}(x^{3})^{2}}\right\} ^{\frac{1}{2}}\ for\ s<0%
\end{array}
\label{g30}
\end{equation}%
Notice that in the identified region (the disc for both sheets) $r=0$ in
both sheets. That is, $r=0$ occurs at $x^{3}=0$ and $s^{2}\leq a^{2}$ for
both sheets $s\gtrless 0$.

The two LCR-congruences $\ell ^{\mu }=\frac{dx_{\ell }^{\mu }}{dr}$ and $%
n^{\mu }=\frac{dx_{n}^{\mu }}{dr}$\ of the flatprint electron LCR-manifold
can be easily implied from the calculations of the previous section. The
starting idea is that the structure coordinates $z^{\alpha }(x)$ provide the
three invariants ($s_{1},s_{2},s_{3}$) along the ray, which\ label the $\ell 
$-ray $x_{\ell }^{\mu }(r)$, and the structure coordinates $z^{\widetilde{%
\alpha }}(x)$ provide the invariants ($s_{1}^{\prime },s_{2}^{\prime
},s_{3}^{\prime }$), which\ label the $n$-ray $x_{n}^{\mu }(r)$. Hence we
simply have the same forms, but we let $r\in (-\infty ,+\infty )$ and at $%
r=0 $ we pass to the second $x_{L}^{\prime \mu }(r),x_{L}^{\prime \mu
}(r)\in 
\mathbb{R}
^{4}$ sheet.

A complete visualization of the rays $w_{L,N}(r;s_{1},s_{2},s_{3})\in U(2)$
taking $r\in (-\infty ,+\infty )$ can be done in the bounded realization of
the flatprint electron (as the $U(2)$ boundary of the $SU(2,2)$\ classical
domain). From the relation 
\begin{equation}
\begin{array}{l}
Y^{0}=\frac{1}{\sqrt{2}}(X^{0}+X^{2})\quad ,\quad Y^{1}=\frac{1}{\sqrt{2}}%
(X^{1}+X^{3}) \\ 
\\ 
Y^{2}=\frac{1}{\sqrt{2}}(X^{0}-X^{2})\quad ,\quad Y^{3}=\frac{1}{\sqrt{2}}%
(X^{1}-X^{3}) \\ 
\end{array}
\label{g31}
\end{equation}%
between the bounded $Y^{ni}$ and unbounded $X^{ni}$ homogeneous coordinates
and%
\begin{equation}
\begin{array}{l}
X^{mi}=%
\begin{pmatrix}
1 & -z^{\widetilde{1}} \\ 
z^{1} & 1 \\ 
-i(z^{0}-ia) & i(z^{\widetilde{0}}-ia)z^{\widetilde{1}} \\ 
-i(z^{0}+ia)z^{1} & -i(z^{\widetilde{0}}+ia)%
\end{pmatrix}
\\ 
\end{array}
\label{g32}
\end{equation}%
we find%
\begin{equation}
\begin{array}{l}
Y^{mi}=\frac{1}{\sqrt{2}}%
\begin{pmatrix}
1-i(z^{0}-ia) & (-1+i(z^{\widetilde{0}}-ia))z^{\widetilde{1}} \\ 
(1-i(z^{0}+ia))z^{1} & 1-i(z^{\widetilde{0}}+ia) \\ 
1+i(z^{0}-ia) & -(1+i(z^{\widetilde{0}}-ia))z^{\widetilde{1}} \\ 
(1+i(z^{0}+ia))z^{1} & 1+i(z^{\widetilde{0}}+ia)%
\end{pmatrix}
\\ 
\end{array}
\label{g33}
\end{equation}%
Like previously, we use the relations (\ref{g18}) to find the labels of $%
L^{\mu }$ rays 
\begin{equation}
\begin{array}{l}
w_{11}=\frac{Y^{21}Y^{12}-Y^{11}Y^{22}}{Y^{01}Y^{12}-Y^{11}Y^{02}}\quad
,\quad w_{12}=\frac{Y^{01}Y^{22}-Y^{21}Y^{02}}{Y^{01}Y^{12}-Y^{11}Y^{02}} \\ 
\\ 
w_{21}=\frac{Y^{31}Y^{12}-Y^{11}Y^{32}}{Y^{01}Y^{12}-Y^{11}Y^{02}}\quad
,\quad w_{12}=\frac{Y^{01}Y^{32}-Y^{31}Y^{02}}{Y^{01}Y^{12}-Y^{11}Y^{02}}%
\end{array}
\label{g34}
\end{equation}%
between the bounded projective $w\in U(2)$ and homogeneous $Y^{ni}$
coordinates, we finally find the rays $w_{L}(r;s_{1},s_{2},s_{3})\in U(2)$
in the complete bounded universe $U(2)$.

The intersection (touching) of the two $%
\mathbb{R}
^{4}$ sheets in $U(2)$ coordinates can be computed by simply making the
Cayley transformation of the cartesian form of the ring singularity. Then we
find that in ($\tau ,\rho ,\sigma ,\chi $) coordinates the ring singularity
(the caustic of the congruence) and its "tube" connecting the two sheets is 
\begin{equation}
\begin{array}{l}
\sigma =\frac{\pi }{2}\quad ,\quad R_{0}^{2}\frac{\sin ^{2}\rho }{(\cos \tau
+\cos \rho )^{2}}\leq a^{2} \\ 
-\pi <\rho <\pi \quad ,\quad -\pi <\tau <\pi%
\end{array}
\label{g35}
\end{equation}%
which apparently contains both rings of the two $%
\mathbb{R}
^{4}$ copies.

In principle we can compute the explicit form of the $L^{\mu }$ ray tracing
in $U(2)$, but it is too complicated. From the cartesian coordinates we have%
\begin{equation}
\begin{array}{l}
x^{0}=\frac{T_{0}\sin \tau }{\cos \tau +\cos \rho } \\ 
x^{1}+ix^{2}=\frac{R_{0}\sin \rho }{\cos \tau +\cos \rho }\sin \sigma \
e^{i\chi }=\sqrt{r^{2}+a^{2}}e^{-i\arctan \frac{a}{r}}\sin \theta
e^{i\varphi } \\ 
x^{3}=\frac{R_{0}\sin \rho }{\cos \tau +\cos \rho }\cos \sigma =r\cos \theta
\\ 
s:=\frac{R_{0}\sin \rho }{\cos \tau +\cos \rho }%
\end{array}
\label{g36}
\end{equation}%
which imply the following relations of the curve determining variables ($%
s_{1},s_{2},s_{3}$) 
\begin{equation}
\begin{array}{l}
s_{1}:=\frac{R_{0}\sin \tau }{\cos \tau +\cos \rho }-r \\ 
s_{2}:=\tan \theta =\frac{r}{\sqrt{r^{2}+a^{2}}}\tan \sigma \\ 
s_{3}:=\varphi =\chi +\arctan \frac{a}{r} \\ 
\\ 
r^{4}-[s^{2}-a^{2}]r^{2}-a^{2}s^{2}\cos ^{2}\sigma =0%
\end{array}
\label{g37}
\end{equation}%
and the convenient affine parameter of the congruence is $r$. The bounded ($%
s_{1},s_{2},s_{3}$) curve in $SU(2)$ for $\tau =0$ can be easily seen.

\section{THE VACUUM}

\setcounter{equation}{0}

The permitted (restricted) holomorphic transformations $z^{\prime
a}=f^{\alpha }(z^{\beta }),\ z^{\prime \widetilde{a}}=f^{\alpha }(z^{%
\widetilde{\beta }})$ may be used\cite{BAOU} to find coordinates (called
regular coordinates) such that (\ref{i6}) take the forms 
\begin{equation}
\begin{array}{l}
\rho _{11}\left( \overline{z^{\alpha }},z^{\widetilde{\alpha }}\right) =%
\func{Im}z^{0}-\phi _{11}(\overline{z^{1}},z^{1},\func{Re}z^{0}) \\ 
\rho _{12}\left( \overline{z^{\alpha }},z^{\widetilde{\alpha }}\right) =z^{%
\widetilde{1}}-\overline{z^{1}}-\phi _{12}(\overline{z^{a}},z^{\widetilde{0}%
}) \\ 
\rho _{22}\left( \overline{z^{\widetilde{\alpha }}},z^{\widetilde{\alpha }%
}\right) =\func{Im}z^{\widetilde{0}}-\phi _{22}(\overline{z^{\widetilde{1}}}%
,z^{\widetilde{1}},\func{Re}z^{\widetilde{0}}) \\ 
\\ 
\phi _{ij}(0)=0\quad ,\quad d\phi _{ij}(0)=0%
\end{array}
\label{v1}
\end{equation}%
where $z^{1},z^{\widetilde{1}}$, are the complex coordinates of $CP^{1}$,
because this regular form of the LCR-structure continues to permit the
following $SL(2,%
\mathbb{C}
)$ transformation%
\begin{equation}
\begin{array}{l}
z^{\prime 1}=\frac{c+dz^{1}}{a+bz^{1}}\quad ,\quad z^{\prime \widetilde{1}}=%
\frac{\overline{c}+\overline{d}z^{\widetilde{1}}}{\overline{a}+\overline{b}%
z^{\widetilde{1}}} \\ 
\\ 
ad-bc=1%
\end{array}
\label{v2}
\end{equation}%
That is, the corresponding spinors transform relative to the conjugate
representations of $SL(2,%
\mathbb{C}
)$%
\begin{equation}
\begin{array}{l}
\begin{pmatrix}
\lambda ^{\prime } \\ 
\lambda ^{\prime }z^{\prime 1}%
\end{pmatrix}%
=%
\begin{pmatrix}
a & b \\ 
c & d%
\end{pmatrix}%
\begin{pmatrix}
\lambda \\ 
\lambda z^{1}%
\end{pmatrix}
\\ 
\\ 
\begin{pmatrix}
-\widetilde{\lambda }^{\prime }z^{\prime \widetilde{1}} \\ 
\widetilde{\lambda }^{\prime }%
\end{pmatrix}%
=%
\begin{pmatrix}
a & b \\ 
c & d%
\end{pmatrix}%
^{\dag -1}%
\begin{pmatrix}
-\widetilde{\lambda }z^{\widetilde{1}} \\ 
\widetilde{\lambda }%
\end{pmatrix}
\\ 
\\ 
ad-bc=1%
\end{array}
\label{v3}
\end{equation}

The action is generally covariant without a precise metric. Therefore the
observed in nature Poincar\'{e} symmetry must be found in the set of
solutions and it must preserve the physical vacuum. The LCR-structure has
been extensively studied\cite{FLAHE1976} in the context of general
relativity under the name of spacetimes with two geodetic and shear free
null congruences. In this context we see that a quite general class of
LCR-manifolds\cite{RAG2013b} take the form of real surfaces of the
grassmannian manifold $G_{4,2}$. The charts of its typical non-homogeneous
coordinates are determined by the invertible pairs of rows. If the first two
rows constitute an invertible matrix, the chart is determined by $\det
\lambda \neq 0$ and the corresponding affine space coordinates $r$ are
defined by 
\begin{equation}
\begin{array}{l}
X=%
\begin{pmatrix}
X^{01} & X^{02} \\ 
X^{11} & X^{12} \\ 
X^{21} & X^{22} \\ 
X^{31} & X^{32}%
\end{pmatrix}%
=\left( 
\begin{array}{c}
\lambda ^{Aj} \\ 
-ir_{A^{\prime }A}\lambda ^{Aj}%
\end{array}%
\right) \\ 
\\ 
r_{A^{\prime }A}=\eta _{ab}r^{a}\sigma _{A^{\prime }A}^{b} \\ 
\end{array}
\label{v4}
\end{equation}%
Then the LCR-structure defining relations take the form 
\begin{equation}
\begin{array}{l}
\rho _{11}(\overline{X^{m1}},X^{n1})=0=\rho _{22}(\overline{X^{m2}},X^{n2})
\\ 
\rho _{12}(\overline{X^{m1}},X^{n2})=0 \\ 
K(X^{mj})=0%
\end{array}
\label{v5}
\end{equation}%
where all the functions are homogeneous relative to $X^{n1}$ and $X^{n2}$\
independently, which must be roots of the homogeneous holomorphic (generally
reducible) Kerr polynomial $K(Z^{m})$. In this context, we see that the
LCR-structures determined by the relations 
\begin{equation}
\begin{array}{l}
X^{mi}E_{mn}X^{nj}=0\quad ,\quad K(X^{mj})=0 \\ 
\\ 
E=%
\begin{pmatrix}
0 & I \\ 
I & 0%
\end{pmatrix}
\\ 
\end{array}
\label{v6}
\end{equation}%
are flat, i.e. they generate a minkowiskian class of metrics $[\eta _{\mu
\nu }]$. Besides, the very fruitful notion of asymptotically flat spacetimes
at null infinity\cite{P-R1984} may be transferred to the asymptotically flat
LCR-structures, which satisfy the conditions 
\begin{equation}
\begin{array}{l}
\overline{X^{m1}}E_{mn}X^{n1}=0=\overline{X^{m2}}E_{mn}X^{n2}\quad ,\quad
\rho _{12}(\overline{X^{m1}},X^{n2})=0 \\ 
\\ 
K(X^{mj})=0 \\ 
\end{array}
\label{v7}
\end{equation}%
Notice that $SU(2,2)$ is the symmetry group of these solutions. The
consideration of open LCR-manifolds implies the removal of a point
(infinity) of the Shilov boundary of the bounded $SU(2,2)$ classical domain%
\cite{PIAT1966}, which restricts the group down to its Poincar\'{e} group up
to an additional dilation group, which will be finally broken by the mass of
the electron. This Poincar\'{e} symmetry group is identified with the
observed Poincar\'{e} symmetry in nature.

Let us now consider the LCR-structures determined by a generally complex
Newman trajectory\cite{NEWM2004} $\xi ^{b}(\tau )$ via the relations 
\begin{equation}
\begin{array}{l}
X=\left( 
\begin{array}{c}
\lambda ^{Aj} \\ 
-ir_{A^{\prime }A}\lambda ^{Aj}%
\end{array}%
\right) =\left( 
\begin{array}{c}
\lambda ^{Aj} \\ 
-i\xi _{A^{\prime }A}\lambda ^{Aj}%
\end{array}%
\right) \\ 
\\ 
\det (r_{A^{\prime }A}-\xi _{A^{\prime }A}(\tau ))=\eta _{ab}(r^{a}-\xi
^{a}(\tau ))(r^{b}-\xi ^{b}(\tau ))=0 \\ 
\\ 
(r_{A^{\prime }A}-\xi _{A^{\prime }A}(\tau _{j}))\lambda ^{Aj}(r)=0 \\ 
\end{array}
\label{v8}
\end{equation}

The coordinate system of an observer is determined by a word line $\xi
^{a}(\tau )=(\tau ,0,0,0)$, which defines a LCR-structure compatible with
the Minkowski metric via the relations 
\begin{equation}
\begin{array}{l}
\begin{pmatrix}
\lambda ^{i} \\ 
-i\xi _{a}(\tau _{i})\sigma ^{a}\lambda ^{i}%
\end{pmatrix}%
=%
\begin{pmatrix}
\lambda ^{i} \\ 
-ix_{a}\sigma ^{a}\lambda ^{i}%
\end{pmatrix}
\\ 
\end{array}
\label{v9}
\end{equation}%
$\tau _{i}$ are the two solutions and $\lambda ^{i}$\ are the spinors of the
corresponding (future pointing) null vectors, i.e. 
\begin{equation}
\begin{array}{l}
\det [(x_{a}-\xi _{a}(\tau ))\sigma ^{a}]=0\quad ,\quad \tau _{1,2}=x^{0}\mp 
\sqrt{(x^{i})^{2}} \\ 
\\ 
(x_{a}-\xi _{a}(\tau _{1}))\sigma _{A^{\prime }A}^{a}\lambda
^{A1}=0=(x_{a}-\xi _{a}(\tau _{2}))\sigma _{A^{\prime }A}^{a}\lambda ^{A2}
\\ 
\end{array}
\label{v10}
\end{equation}%
and the corresponding null vectors are\ 
\begin{equation}
\begin{array}{l}
\begin{pmatrix}
\sqrt{(x^{i})^{2}} \\ 
x^{1} \\ 
x^{2} \\ 
x^{3}%
\end{pmatrix}%
=\lambda ^{1\dag }\sigma ^{a}\lambda ^{1}\quad ,\quad 
\begin{pmatrix}
\sqrt{(x^{i})^{2}} \\ 
-x^{1} \\ 
-x^{2} \\ 
-x^{3}%
\end{pmatrix}%
=\lambda ^{2\dag }\sigma ^{a}\lambda ^{2} \\ 
\end{array}
\label{v11}
\end{equation}%
This shows that the two spinors $\lambda ^{A1}$ and $\lambda ^{A2}$\ define
spatially inverted null vectors. Hence, these two spinors must belong to the
conjugate chiral representations of the $SL(2,%
\mathbb{C}
)$ group. This means that the spinors defined by the left and right columns
of the homogeneous coordinates of the vacuum (degenerate) LCR-structure must
have\cite{GMS} opposite chiralities, because parity is an external
automorphism of the orthochronous proper Lorentz group. It corresponds a
spinor of the fundamental representation to a spinor of its conjugate
representation.

Note that this trajectory satisfies the Poincar\'{e} invariant normalization
condition $\eta _{ab}\frac{d\xi ^{a}}{d\tau }\frac{d\xi ^{b}}{d\tau }=1$.
This is the vacuum of the precise observer. Any other Poincar\'{e}
transformed observer $\xi ^{a}(\tau )=(v^{0}\tau ,v^{i}\tau +c^{i})$ has a
pseudo-conformally equivalent vacuum (LCR-structure). In [paper I], I have
already shown that this vacuum is invariant under the Poincar\'{e}
transformations determined with infinity fixed with the projective chart
condition $\det \lambda =0$.

In the same chart we may define a different LCR-manifold, which apparently
belongs to a different representation of the Poincar\'{e} group, because it
is determined by the two real trajectories $\xi _{(\mp )}^{a}(\tau )=(\tau
,0,0,\mp \tau )$, which satisfy the Poincar\'{e} invariant normalization
condition $\eta _{ab}\frac{d\xi ^{a}}{d\tau }\frac{d\xi ^{b}}{d\tau }=0$.
The homogeneous coordinates of this LCR-structure and the appropriate
structure coordinates $z^{\alpha },z^{\widetilde{\beta }}$ are 
\begin{equation}
\begin{array}{l}
X=%
\begin{pmatrix}
1 & 0 \\ 
0 & 1 \\ 
-i(x^{0}-x^{3}) & i(x^{1}-ix^{2}) \\ 
i(x^{1}+ix^{2}) & -i(x^{0}+x^{3})%
\end{pmatrix}%
=%
\begin{pmatrix}
1 & 0 \\ 
0 & 1 \\ 
-iz^{0} & -iz^{\widetilde{1}} \\ 
-iz^{1} & -iz^{\widetilde{0}}%
\end{pmatrix}
\\ 
\end{array}
\label{v12}
\end{equation}%
In complete analogy to the proceeding vacuum [paper I], the degenerate
relations 
\begin{equation}
\begin{array}{l}
z^{0}-\overline{z^{0}}=0\quad ,\quad z^{\widetilde{0}}-\overline{z^{%
\widetilde{0}}}=0\quad ,\quad z^{\widetilde{1}}-\overline{z^{1}}=0 \\ 
\end{array}
\label{v13}
\end{equation}%
remain formally invariant under the Lorentz subgroup 
\begin{equation}
\begin{array}{l}
\begin{pmatrix}
X_{1}^{\prime i} \\ 
X_{2}^{\prime i}%
\end{pmatrix}%
=%
\begin{pmatrix}
B_{11}^{i} & 0 \\ 
0 & B_{22}^{i}%
\end{pmatrix}%
\left( 
\begin{array}{c}
X_{1}^{i} \\ 
X_{2}^{i}%
\end{array}%
\right) \\ 
\\ 
B_{11}^{1}=%
\begin{pmatrix}
a & b \\ 
c & d%
\end{pmatrix}%
\quad ,\quad B_{11}^{2}=%
\begin{pmatrix}
\overline{d} & -\overline{c} \\ 
-\overline{b} & \overline{a}%
\end{pmatrix}
\\ 
\\ 
B_{11}^{i\dagger }B_{22}^{i}=I \\ 
\end{array}
\label{v14}
\end{equation}%
and the real translation subgroup%
\begin{equation}
\begin{array}{l}
\begin{pmatrix}
X_{1}^{\prime } \\ 
X_{2}^{\prime }%
\end{pmatrix}%
=%
\begin{pmatrix}
I & 0 \\ 
B_{21} & I%
\end{pmatrix}%
\left( 
\begin{array}{c}
X_{1} \\ 
X_{2}%
\end{array}%
\right) \\ 
\\ 
B_{21}+B_{21}^{\dagger }=0 \\ 
\end{array}
\label{v15}
\end{equation}%
of the $SU(2,2)$ group.

In the following sections, quantum field theory of the standard model will
be viewed as a harmonic analysis of distribution valued free fields rigged
Hilbert-Fock space of the Poincar\'{e} group representations. This
mathematical picture is based on our consideration that the universe is
described by a LCR-structure condition (\ref{v1}) $\rho _{ij}=0$. The
singular regions of the Schwartz distributions determine the "location" of
the particles and their potentials (gravitational, electroweak and Higgs)
are their corresponding representative functions. Recall that these
representatives have locally integrable singularities at the location of the
particle and are regular outside. This framework is characterized by the
geometric Kaehler structure \ 
\begin{equation}
\begin{array}{l}
ds^{2}=2\frac{\partial ^{2}\det (\rho _{ij})}{\partial z^{a}\partial 
\overline{z^{b}}}dz^{a}d\overline{z^{b}}\quad ,\quad \omega =2i\frac{%
\partial ^{2}\det (\rho _{ij})}{\partial z^{a}\partial \overline{z^{b}}}%
dz^{a}\wedge d\overline{z^{b}} \\ 
\end{array}
\label{v16}
\end{equation}%
This means that the universe LCR-submanifold $\rho _{ij}=0$ is real analytic
in the "empty" space. Besides, it is a totally real lagrangian submanifold
of $%
\mathbb{C}
^{4}$ with the induced metric identified with the Einstein metric of the
universe. The reader must realize that here we are not talking about the
phenomenological astrophysical metrics, but for everything from the very
"small" to the "largest" possible. We know \cite{BAOU} that any totally real
submanifold admits an analytic transformation $z^{b}=f_{p}^{b}(r^{a})$ in
the neighborhoods of their analytic points $p$, which trivializes the real
submanifold \ 
\begin{equation}
\begin{array}{l}
\rho _{ij}=\frac{\widehat{r}-\widehat{r}^{\dag }}{2i}=0 \\ 
\\ 
\widehat{r}:=%
\begin{pmatrix}
r^{0}-r^{3} & -(r^{1}-ir^{2}) \\ 
-(r^{1}+ir^{2}) & r^{0}+r^{3}%
\end{pmatrix}
\\ 
r^{a}:=x^{a}+iy^{a}%
\end{array}
\label{v17}
\end{equation}%
This analytic transformation \textbf{does not} preserve the LCR-structure,
but it constitutes a legitimate transformation in the Kaehler manifold and
the real variables are the well-known Darboux real coordinates. The metric
and the symplectic 2-form in $%
\mathbb{C}
^{4}$ are \ 
\begin{equation}
\begin{array}{l}
ds^{2}=\frac{1}{2}\tsum\limits_{a,b}\frac{\partial ^{2}(-(\overline{r^{c}}%
-r^{c})^{2})}{\partial r^{a}\partial \overline{r^{b}}}dr^{a}d\overline{r^{b}}%
=\eta _{ab}dr^{a}d\overline{r^{b}}=\eta _{ab}(dx^{a}dx^{b}+dy^{a}dy^{b}) \\ 
\omega =\frac{i}{2}\tsum\limits_{a,b}\frac{\partial ^{2}(-(\overline{r^{c}}%
-r^{c})^{2})}{\partial r^{a}\partial \overline{r^{b}}}dr^{a}\wedge d%
\overline{r^{b}}=i\eta _{ab}dr^{a}\wedge d\overline{r^{b}}=2\eta
_{ab}dx^{a}\wedge dy^{b}%
\end{array}
\label{v18}
\end{equation}%
Apparently this locally analytic transformation \textbf{cannot be} extended
to the singular regions of the universe. Therefore our approach has to be
properly adapted. We will keep the Minkowski coordinates and we will extend
the set of functions to the set of Schwartz distributions (generalized
functions).

Notice that the 2-form (\ref{v18}) is not the ordinary form of the
symplectic phase space. The emergence of the Minkowski metric implies that
locally the phase space is the $SU(2,2)$ symmetric Siegel domain $\frac{%
\widehat{r}-\widehat{r}^{\dag }}{2i}>0$ with the corresponding projective
form $X^{\dag }E_{U}X>0$. The (singular) Schwartz distributions are
characterized by their characteristic locally integrable functions, which we
will call potentials. Their derivatives \textbf{are not} locally integrable,
but they are proper distributions because of the space of the appropriate
test functions. We will use the Bogoliubov perurbative approach\cite{BOG1975}
based on the tempered distributions and the corresponding rigged Hilbert
space of the Poicar\'{e} representations (free fields).

In order to clarify the subsequent mathematical approach let me mention the
well known mathematical problem that Pythagora affronted and the final
solution that Cauchy gave about 2000 years later. Integers and their ratios
(the rational numbers) were the only numbers that Pythagora new. Pythagorian
theorem gave him the possibility to compute/measure square roots of numbers
(like $\sqrt{2}$). Near the end of his life, he realized that $\sqrt{2}$\ is
not a rational number. We now know that this is a real number, and we need
an infinite series of rational numbers to approximate it. It was Gelfand who
through his triple (rigged Hilbert space) permitted the harmonic expansion
of the generalized functions to representations of the Poincar\'{e} group.
Hence the discovery of the electron and neutrino as free stable
distributional solitons, the computation of its generalized state,
gravitational and electroweak potentials turn out to be a mathematical and
not a physical problem. For the computation of the hypotenuse we simply need
the proper completion of the rational numbers to the complete set (Hilbert
space) of real numbers.

\section{BOGOLIUBOV'S\ PERTURBATIVE QFT}

\setcounter{equation}{0}

The Bogoliubov-Medvedev-Polivanov method\cite{BOG1975} approaches the
axiomatic formulation of a quantum field theory starting from the S-matrix
and the introduction of a "switching on and off" function $c(x)\in \lbrack
0,1]$ and assuming the following expansion of the S-matrix \ 
\begin{equation}
\begin{array}{l}
S=1+\underset{n\geq 1}{\sum }\frac{1}{n}\int
S_{n}(x_{1},x_{2}...x_{n})c(x_{1})c(x_{2})...c(x_{n})[dx] \\ 
\end{array}
\label{e0}
\end{equation}%
where $S_{n}(x_{1},x_{2}...x_{n})$ depends on the complete free field
functions (the local Poincar\'{e} representations of the particles) and not
its separate "positive" and "negative" frequency parts. That is, the
S-matrix is an operator in the Fock space of free relativistic particles. It
satisfies the following axioms \ 
\begin{equation}
\begin{array}{l}
Poincar\acute{e}\ covariance:\quad
U_{P}S_{n}(x_{1},x_{2}...x_{n})U_{P}^{\dag }=S_{n}(Px_{1},Px_{2}...Px_{n})
\\ 
Unitarity:\quad SS^{\dag }=S^{\dag }S=1 \\ 
Microcausality:\quad \frac{\delta }{\delta c(x)}[\frac{\delta S(c)}{\delta
c(x)}S^{\dag }(c)]=0\quad for\quad x\precsim y \\ 
Correspondance\ principle:\quad S_{1}(x)=iL_{int}[\phi (x)] \\ 
\end{array}
\label{e0a}
\end{equation}%
where $\phi (x)$ denotes the free particle fields and $x\precsim y$ means $%
x^{0}<y^{0}$ or $(x-y)^{2}<0$. A general solution of these conditions is the
Dyson form of the time evolution unitary matrix (S-matrix) \ 
\begin{equation}
\begin{array}{l}
S=T[\exp (i\mathbf{L}[\phi (x);c(x))] \\ 
\\ 
\mathbf{L}[\phi (x);c(x)]=L_{Int}[\phi (x)]c(x)+\underset{n\geq 1}{\sum }%
\frac{1}{n}\int \Lambda _{n+1}(x,x_{1}...x_{n})c(x)c(x_{1})...c(x_{n})[dx]
\\ 
\end{array}
\label{e0b}
\end{equation}%
where $\Lambda _{n}(x,x_{1}...x_{n})$ are quasilocal quantities, which
permit the renormalization process. This order by order construction of a
finite S-matrix (with possibly infinite hamiltonian and lagrangian) provides
a well established algorithm to distinguish renormalizable with
non-renormalizable interaction lagrangians\cite{BOG1980}.

The advantage of the BMP procedure is that it can be used in the opposite
sense. Knowing the Poincar\'{e} representations, they are identified with
"free particles" with precise mass and spin. Then they are described with
the corresponding free fields, which are used to write down an effective
interaction lagrangian, suggested by the fundamental dynamics. In the
present case, the fundamental dynamics is the PCFT and the particles are the
solitonic solutions and their corresponding potentials which satisfy the
wave equations. The suggested interaction takes the place of the
"correspondence principle" in the BMP procedure. The order by order
computation introduces counterterms to the action (with up to first order
derivatives). If the number of the forms of the counterterms is finite, the
action is normalizable and the model is considered compatible with quantum
mechanics.

Epstein-Glaser\cite{EpGl} noticed that the renormalization procedure is an
artifact of the improper replacement of the Dyson time-ordering with the
step function (distribution). The multiplication of this distribution with
the Wightman distributions \textbf{is not} mathematically permitted. They
developed a proper regularization procedure using the scale function. This
is very well described in the first book of Scharf\cite{Sch1}. In his second
book\cite{Sch2}, Scharf used the operational Krein structure to annihilate
order by order the non-physical modes of the free fields. These two
ameliorations made the Bogoliubov procedure an algorithmic Poincar\'{e}
group harmonic expansion of the S-matrix as an operator distribution valued
in the Gelfand rigged Hilbert-Fock space\cite{BOG1975} 
\begin{equation}
\begin{array}{l}
\mathcal{S}(%
\mathbb{R}
^{n})\subset L^{2}(%
\mathbb{R}
^{n})\subset \mathcal{S}^{\prime }(%
\mathbb{R}
^{n}) \\ 
\end{array}
\label{e0c}
\end{equation}%
of the tempered distributions, where the space $\mathcal{S}(%
\mathbb{R}
^{n})$ is dense in $L^{2}(%
\mathbb{R}
^{n})$ relative to the supremum definition topology. Hence we only need now
is to found the asymptotic fields (particles). The existence of the electron
LCR-structure with its gravitational and electromagnetic potentials
"triggers" the algorithmic process, providing the entire S-matrix of quantum
electrodynamics.

The Kerr-Newman electrified spacetime is one of the physically interesting
solutions in general relativity. It admits two geodetic and shear free null
congruences, which are related with the Kerr polynomial (\ref{g10a}). It
also admits two commuting killing vectors, which are identified with the
time-translation and $z$-rotation generators of the Poincar\'{e} group.
Carter's\cite{CART1968} discovery that the gyromagnetic ratio of the
Kerr-Newman manifold is fermionic (that of the electron $g=2$)\cite{N-W1974}
shocked the community of general relativists. Many tried to identify the
Kerr-Newman spacetime with the electron without success.

After the identification of the phenomenological Poincar\'{e} symmetry with
the $SU(2,2)$ subgroup, which preserves infinity, it is straight-forward to
compute the asymptotically flat LCR-structure, which admits the
time-translation and $z$-rotation Killing vectors. It coincides with the
LCR-structure found applying the Kerr-Schild ansatz procedure\cite{RAG1999}.
Recall that the electron is the unique charged stable leptonic particle of
current phenomenology.

In the linearized Einstein-gravity approximation, the Kerr-Schild ansatz
coincides with the approximation itself. This fact facilitates our
calculation and interpretation. Hence, the $G_{4,2}$ point of the flatprint
of the electron LCR-structure is determined from the static trajectory $\xi
^{b}=(\tau ,0,0,ia)$. The corresponding two spinors $\lambda ^{Ai}$, which
appear in its representation in the homogeneous coordinates have the form \ 
\begin{equation}
\begin{array}{l}
\lambda ^{Ai}=%
\begin{pmatrix}
x^{1}-ix^{2} & x^{1}-ix^{2} \\ 
x^{0}-x^{3}-\tau _{1}-ia & x^{0}-x^{3}-\tau _{2}-ia%
\end{pmatrix}
\\ 
\\ 
\tau _{1,2}=x^{0}\mp \sqrt{(x^{1})^{2}+(x^{2})^{2}+(x^{3}-ia)^{2}} \\ 
\end{array}
\label{e1}
\end{equation}%
and the flat null tetrad is%
\begin{equation}
\begin{array}{l}
L^{a}=\frac{1}{\sqrt{2}}\overline{\lambda }^{A^{\prime }1}\lambda
^{B1}\sigma _{A^{\prime }B}^{a}\quad ,\quad N^{a}=\frac{1}{\sqrt{2}}%
\overline{\lambda }^{A^{\prime }2}\lambda ^{B2}\sigma _{A^{\prime }B}^{a} \\ 
\\ 
M^{a}=\frac{1}{\sqrt{2}}\overline{\lambda }^{A^{\prime }2}\lambda
^{B1}\sigma _{A^{\prime }B}^{a} \\ 
\\ 
\epsilon _{AB}\lambda ^{A1}\lambda ^{B2}=1 \\ 
\end{array}
\label{e2}
\end{equation}%
where the spinors have been properly normalized. In the Lindquist
coordinates it takes the form%
\begin{equation}
\begin{array}{l}
L^{\mu }\partial _{\mu }=\partial _{t}+\partial _{r} \\ 
\\ 
N^{\mu }\partial _{\mu }=\frac{r^{2}+a^{2}}{2(r^{2}+a^{2}\cos ^{2}\theta )}%
\left( \partial _{t}-\partial _{r}+\frac{2a}{r^{2}+a^{2}}\partial _{\varphi
}\right) \\ 
\\ 
M^{\mu }\partial _{\mu }=\frac{1}{\sqrt{2}(r+ia\cos \theta )}\left( ia\sin
\theta \partial _{t}+\partial _{\theta }+\frac{i}{\sin \theta }\partial
_{\varphi }\right)%
\end{array}
\label{e2a}
\end{equation}%
Its covariant form is%
\begin{equation}
\begin{array}{l}
L_{\mu }dx^{\mu }=dt-dr-a\sin ^{2}\theta \ d\varphi \\ 
\\ 
N_{\mu }dx^{\mu }=\frac{r^{2}+a^{2}}{2(r^{2}+a^{2}\cos ^{2}\theta )}[dt+%
\frac{r^{2}+2a^{2}\cos ^{2}\theta -a^{2}}{r^{2}+a^{2}}dr-a\sin ^{2}\theta \
d\varphi ] \\ 
\\ 
M_{\mu }dx^{\mu }=\frac{-1}{\sqrt{2}(r+ia\cos \theta )}[-ia\sin \theta \
(dt-dr)+(r^{2}+a^{2}\cos ^{2}\theta )d\theta + \\ 
\qquad \qquad +i\sin \theta (r^{2}+a^{2})d\varphi ]%
\end{array}
\label{e3}
\end{equation}%
The Kerr-Schild ansatz gives\cite{RAG1990}$^{,}$\cite{RAG1999} the general
form of the curved LCR-manifold\ 
\begin{equation}
\begin{array}{l}
\ell _{\mu }=L_{\mu }\quad ,\quad m_{\mu }=M_{\mu }\quad ,\quad n_{\mu
}=N_{\mu }+\frac{h(r)}{2(r^{2}+a^{2}\cos ^{2}\theta )}\ L_{\mu } \\ 
\end{array}
\label{e4}
\end{equation}%
where $h(r)$ is an arbitrary function. Notice that for $h(r)=-2mr+e^{2}$ the
Kerr-Newman space-time is found with electromagnetic potential\ 
\begin{equation}
\begin{array}{l}
A=\frac{qr}{4\pi (r^{2}+a^{2}\cos ^{2}\theta )}(dt-dr-a\sin ^{2}\theta
d\varphi ) \\ 
\\ 
r^{4}-[(x^{1})^{2}+(x^{2})^{2}+(x^{3})^{2}-a^{2}]r^{2}-a^{2}(x^{3})^{2}=0%
\end{array}
\label{e4a}
\end{equation}%
The fermionic parameter $a$ appears in the electromagnetic and gravitational
potentials (dressings) of the electron. Besides the electromagnetic
potential is proportional to the null tetrad covector $\ell _{\mu }$. The
explicit computation of the electroweak connection (in section V) will make
clear that it is not a computational accident.

From the deriving relations \ 
\begin{equation}
\begin{array}{l}
(x_{a}-\xi _{a}(\tau _{j}))\sigma _{A^{\prime }A}^{a}\lambda ^{Aj}=0 \\ 
\\ 
x^{a}-\xi ^{a}(\tau _{j})=%
\begin{pmatrix}
\pm \sqrt{(x^{1})^{2}+(x^{2})^{2}+(x^{3}-ia)^{2}} \\ 
x^{1} \\ 
x^{2} \\ 
x^{3}-ia%
\end{pmatrix}%
\end{array}
\label{e5}
\end{equation}%
we see that $\lambda ^{A2}$ satisfies a relation implied after a temporal
reflection of the corresponding relation that $\lambda ^{A1}$ satisfies,
because $(x^{0}-\tau _{2})=-(x^{0}-\tau _{1})$. Hence, these two spinors
must belong to the conjugate chiral representations of the $SL(2,%
\mathbb{C}
)$ group. This means that the spinors defined by the left and right columns
of the homogeneous coordinates of the electron LCR-structure must have\cite%
{GMS} opposite chiralities, because temporal reflection (like parity) is an
external automorphism of the orthochronous proper Lorentz group. I want to
point out that this relation is generalized only in the case of
LCR-structures determined by one trajectory. In the general case of left and
right columns of $X^{ni}$ implied by different trajectories (reduced Kerr
polynomials), they are not related with such a discrete symmetry. One has to
go back to the regular LCR-structure coordinates (\ref{v1}) to reveal
opposite chirality between left and right columns of the homogeneous
coordinates of $G_{4,2}$.

Because of the importance of the chirality emergence, I will now explicitly
show that the temporal (and spatial) reflection, applied directly to the
geodetic and shear free condition on $\lambda ^{Aj}(x)$ 
\begin{equation}
\begin{array}{l}
\lambda ^{A}\lambda _{B}\sigma _{A^{\prime }A}^{b}\frac{\partial }{\partial
x^{b}}\lambda ^{B}=0\quad ,\quad \lambda ^{A}=\lambda ^{0}%
\begin{pmatrix}
1 \\ 
\lambda%
\end{pmatrix}
\\ 
\multicolumn{1}{c}{\Updownarrow} \\ 
\frac{\partial \lambda }{\partial x^{0^{\prime }0}}+\lambda \frac{\partial
\lambda }{\partial x^{0^{\prime }1}}=0\ \ \quad and\ \quad \ \frac{\partial
\lambda }{\partial x^{1^{\prime }0}}+\lambda \frac{\partial \lambda }{%
\partial x^{1^{\prime }1}}=0 \\ 
\end{array}
\label{e6}
\end{equation}%
implies the change of the $SL(2,%
\mathbb{C}
)$ representation.

Using my notation 
\begin{equation}
\begin{array}{l}
x_{A^{\prime }A}=x_{\mu }\sigma _{A^{\prime }A}^{\mu }=%
\begin{pmatrix}
x^{0}-x^{3} & -(x^{1}-ix^{2}) \\ 
-(x^{1}+ix^{2}) & x^{0}+x^{3}%
\end{pmatrix}
\\ 
\\ 
x^{A^{\prime }A}=x^{\mu }\sigma _{\mu }^{A^{\prime }A}=%
\begin{pmatrix}
x^{0}+x^{3} & (x^{1}+ix^{2}) \\ 
(x^{1}-ix^{2}) & x^{0}-x^{3}%
\end{pmatrix}
\\ 
\end{array}
\label{e7}
\end{equation}%
I make the temporal reflection 
\begin{equation}
\begin{array}{l}
x^{\prime }=%
\begin{pmatrix}
-x^{0}-x^{3} & -(x^{1}-ix^{2}) \\ 
-(x^{1}+ix^{2}) & -x^{0}+x^{3}%
\end{pmatrix}%
=-\epsilon \overline{x}\epsilon ^{-1} \\ 
\\ 
\epsilon =%
\begin{pmatrix}
0 & 1 \\ 
-1 & 0%
\end{pmatrix}
\\ 
\end{array}
\label{e8}
\end{equation}%
which implies 
\begin{equation}
\begin{array}{l}
\frac{\partial \lambda ^{\prime }}{\partial x^{\prime 0^{\prime }0}}+\lambda
^{\prime }\frac{\partial \lambda ^{\prime }}{\partial x^{\prime 0^{\prime }1}%
}=0\ \ \quad and\ \quad \ \frac{\partial \lambda ^{\prime }}{\partial
x^{\prime 1^{\prime }0}}+\lambda ^{\prime }\frac{\partial \lambda ^{\prime }%
}{\partial x^{\prime 1^{\prime }1}}=0 \\ 
\multicolumn{1}{c}{\Updownarrow} \\ 
\lambda ^{\prime }=\frac{-1}{\overline{\lambda }}\quad ,\quad \lambda
^{\prime A}=\lambda ^{\prime 0}%
\begin{pmatrix}
1 \\ 
\lambda ^{\prime }%
\end{pmatrix}%
=-\frac{\lambda ^{\prime 0}}{\overline{\lambda }}%
\begin{pmatrix}
-\overline{\lambda } \\ 
1%
\end{pmatrix}
\\ 
\multicolumn{1}{c}{\Updownarrow} \\ 
\frac{\partial \lambda }{\partial x^{0^{\prime }0}}+\lambda \frac{\partial
\lambda }{\partial x^{0^{\prime }1}}=0\ \ \quad and\ \quad \ \frac{\partial
\lambda }{\partial x^{1^{\prime }0}}+\lambda \frac{\partial \lambda }{%
\partial x^{1^{\prime }1}}=0%
\end{array}
\label{e9}
\end{equation}%
As expected\cite{GMS}, the representation of the spinor changes to its
conjugate one.

We saw that the chirality distinction is fundamental in the pseudo-conformal
field theory (PCFT). The massive Poincar\'{e} representation of the flat
LCR-structure is determined with the complex linear trajectory%
\begin{equation}
\begin{array}{l}
\xi ^{b}(s)=v^{b}s+c^{b}+ia^{b}\quad ,\quad \overset{.}{(\xi ^{b}}%
)^{2}=(v^{b})^{2}=1 \\ 
\end{array}
\label{e10}
\end{equation}%
where $v^{b},c^{b},a^{b}$\ are the real constants, which represent the
constant velocity, the initial position and the spin of the classical
configuration of the electron. Note that the present normalization of the
parametrization is properly changed in order to assure the massive character
of the representation. In the next section we will argue that the complex
linear trajectories with $\overset{.}{(\xi ^{b}})^{2}=0$ is related to the
neutrino. That is the neutrino is the "boosted" electron in the projective
LCR-structure formalism. Considering the accelerating electron as a ruled
surface of $CP^{3}$ with a general trajectory $\xi ^{b}(\tau )$ the electron
corresponds to the non-vanishing gaussian curvature and its neutrino to the
corresponding tangential developable surface, which ha vanishing gaussian
curvature. Hence my conclusion is that the electron and its massless
neutrino have to be treated as free Dirac fields in the Bogoliubov
expansion. Besides recall that the computed\cite{CART1968} gyromagnetic
ratio of the electron LCR-manifold is fermionic $g=2$.

\subsection{Derivation of quantum electrodynamics}

In addition to the symmetric tensor $g_{\mu \nu }$, the LCR-structure tetrad
also defines a class of three self-dual and three antiself-dual 2-forms
(relative to the defined metric (\ref{i10}))

\begin{equation}
\begin{array}{l}
V^{0}=\ell \wedge m\quad ,\quad V^{\widetilde{0}}=n\wedge \overline{m}\quad
,\quad V=2\ell \wedge n-2m\wedge \overline{m} \\ 
\end{array}
\label{e15}
\end{equation}%
which satisfy the relations

\begin{equation}
\begin{array}{l}
dV^{0}=[(2\varepsilon -\rho )n+(\tau -2\beta )\overline{m}]\wedge V^{0} \\ 
\\ 
dV^{\widetilde{0}}=[(\mu -2\gamma )\ell +(2\alpha -\pi )m]\wedge V^{%
\widetilde{0}} \\ 
\\ 
dV=[2\mu \ell -2\rho n-2\pi m+2\tau \overline{m}]\wedge V \\ 
\end{array}
\label{e16}
\end{equation}%
where the small greek letters are the connection parameters of the
spin-coefficient formalism\cite{P-R1984}. In fact any non conformally flat
metric, which admits geodetic and shear free null directions, define a
finite number of triplets of such self-dual 2-forms. This number is related
to the Petrov type of the metric, and we will discuss it below.

If the LCR-structure is realizable\cite{BAOU}, there are always functions
such that

\begin{equation}
\begin{array}{l}
0=d(dz^{0}\wedge dz^{1})=d[(f_{0}^{0}f_{1}^{1}-f_{0}^{1}f_{1}^{0})\ell
\wedge m] \\ 
\\ 
0=d(dz^{\widetilde{0}}\wedge dz^{\widetilde{1}})=d[(f_{\widetilde{0}}^{%
\widetilde{0}}f_{\widetilde{1}}^{\widetilde{1}}-f_{\widetilde{0}}^{%
\widetilde{1}}f_{\widetilde{1}}^{\widetilde{0}})n\wedge \overline{m} \\ 
\end{array}
\label{e17}
\end{equation}%
But for the third self-dual 2-form $V$, there is not always a function,
which makes it closed i.e. such that $d(fV)=0$. This happens if

\begin{equation}
\begin{array}{l}
d[2\mu \ell -2\rho n-2\pi m+2\tau \overline{m}]=0 \\ 
\end{array}
\label{e18}
\end{equation}%
In fact, if there is a member of the tetrad-Weyl equivalent class of
2-forms, which implies (\ref{e18}), this member may be assumed as the
physical representative, because it defines a conserved "charge". That is,
the existence of a 2-form which defines the conserved quantity "charge"
breaks the tetrad-Weyl symmetry down to the ordinary Weyl symmetry. The
remaining Weyl symmetry will be finally restricted to one tetrad, from the
definition of the mass from the Einstein gravity source.

The electron LCR-structure (\ref{e3}) satisfies this condition, because

\begin{equation}
\begin{array}{l}
2\mu \ell -2\rho n-2\pi m+2\tau \overline{m}=d[\ln (r-ia\cos \theta )^{2}]
\\ 
\end{array}
\label{e19}
\end{equation}%
Hence, the self-dual 2-form

\begin{equation}
\begin{array}{l}
F^{+}=\frac{1}{(r-ia\cos \theta )^{2}}(2\ell \wedge n-2m\wedge \overline{m}%
)=F-i\ \ast F \\ 
\\ 
A=\frac{qr^{3}}{4\pi (r^{4}+a^{2}(x^{3})^{2})}(dx^{0}-\frac{rx^{1}-ax^{2}}{%
r^{2}+a^{2}}dx^{1}-\frac{rx^{2}+ax^{1}}{r^{2}+a^{2}}dx^{2}-\frac{x^{3}}{r}%
dx^{3})%
\end{array}
\label{e20}
\end{equation}%
is closed. It defines a real 2-form $F$, which is identified with the
electromagnetic field, as we see from the its electromagnetic potential,
written in cartesian coordinates in order to compare it with the first order
electron dressing computed below from the Bogoliubov procedure.

The solitonic feature of the electron LCR-structure is protected by the
non-vanishing of all the three relative invariants $\Phi _{i}$ of the
LCR-structure. Recall that the trivial vacuum has all the three relative
invariants equal to zero. In the physical interpretation, this LCR-manifold
has to be identified with the electron and its complex conjugate structure
with the positron, because it has opposite charge.

Notice that the electromagnetic field is essentially determined by the
flatprint of the electron LCR-structure. The second term of the tetrad form
of the Kerr-Schild ansatz (\ref{e4}) does not contribute to the definition
of $F$. Therefore we should consider that the electromagnetic field is a
particle (Poincar\'{e} representation determined by the singularity of the
LCR-structure). Hence the correct field equation is

\begin{equation}
\begin{array}{l}
d\ast F=j\quad ,\quad dj=0 \\ 
\\ 
\partial _{\mu }F^{\mu \nu }=j^{\nu }\quad ,\quad \partial _{\nu }j^{\nu }=0%
\end{array}
\label{e21}
\end{equation}%
where the singularity gives a conserved current. The implied conserved
quantity is the electron charge.

The positron is identified with the conjugate electron LCR-structure, which
is found by simply interchanging ($m\Leftrightarrow \overline{m})$. Then the
metric remains the same, which implies that electron and positron have the
same masses. But the 2-forms change, implying that electron and positron
have opposite charges.

The energy-momentum are the conserved quantities determined from the source $%
T_{(p)}^{\mu \nu }$ of the (linearized) Einstein equation. This satisfies
the Bianchi identities, which must be valid even at the "singularities".
Recall that this point was essentially used by Einstein and coworkers\cite%
{E-I-H1938} to derive the equations of motion. On the other hand the
EM-equations happens to be satisfied by the static soliton. They are not
satisfied for any LCR-structure.

I want to point out that the definition of the LCR-structure (\ref{i5})
implies the tetrad-Weyl invariants $F_{i}=dZ_{i}$ and the relative
invariants $\Phi _{i}$. In the case of the Kerr-Schild ansatz (\ref{e4}) the
invariants of the LCR-structure \ 
\begin{equation}
\begin{array}{l}
F_{1}=dZ_{1}=\frac{4ra^{2}\sin \theta \cos \theta }{(r^{2}+a^{2}\cos \theta
)^{2}}dr\wedge d\theta \\ 
\\ 
F_{2}=dZ_{2}=\frac{4ra^{2}\sin \theta \cos \theta }{(r^{2}+a^{2}\cos \theta
)^{2}}dr\wedge d\theta \\ 
\\ 
F_{3}=dZ_{3}=-\frac{4ra^{2}\sin \theta \cos \theta }{(r^{2}+a^{2}\cos \theta
)^{2}}dr\wedge d\theta \\ 
\end{array}
\label{e24}
\end{equation}%
do not depend on $h(r)$, and the relative invariants 
\begin{equation}
\begin{tabular}{l}
$\rho -\overline{\rho }=\frac{-2ia\cos \theta }{(r+ia\cos \theta )(r-ia\cos
\theta )}=i\Phi _{1}$ \\ 
$\mu -\overline{\mu }=\frac{ia(r^{2}+a^{2}+h)\cos \theta }{(r+ia\cos \theta
)^{2}(r-ia\cos \theta )^{2}}=-i\Phi _{2}$ \\ 
$\tau +\overline{\pi }=\frac{i\sqrt{2}ar\sin \theta }{(r+ia\cos \theta
)^{2}(r-ia\cos \theta )}=\Phi _{3}$%
\end{tabular}
\label{e25}
\end{equation}%
do not vanish. Hence, $h(r)$ is not obstructed from taking the
self-consistent form.

Summarizing, I have already shown that the static LCR-manifold is a soliton
protected by its mass and its relative invariants, and it belongs to the
massive spinorial representation. Hence, it is represented with the Dirac
free field $\psi (x)$, which satisfies the massive Dirac equation. This
Dirac field represents the "left" and "right" columns $X^{nj}$ of the
homogeneous grassmannian coordinates of the moving electron-soliton. It has
an EM-potential (dressing), which satisfies the massless wave equation,
hence it is a spin-1 particle represented with a vector field $A_{\mu }(x)$.
The interaction of these two formal "quantum" free fields is apparently the
well known electromagnetic interaction 
\begin{equation}
\begin{array}{l}
L_{EM}=e\overline{\psi _{e}}\gamma ^{\mu }\psi _{e}A_{\mu } \\ 
\end{array}
\label{e28}
\end{equation}

The Bogoliubov procedure permit us to find the 1st order electromagnetic
potential

\begin{equation}
\begin{array}{l}
A_{1_{\mu }}(x;1)\simeq \frac{-i}{2}\overset{\ast }{\Phi }_{1}\frac{\delta 
\widehat{S}_{2}(J)}{\delta J^{\mu }(x)}\Phi _{1}|_{J^{\mu }=0} \\ 
\\ 
\widehat{S}_{2}(J)=\tint T((L_{I}(x_{1})+A_{\nu }(x_{1})J^{\nu
}(x_{1}))(L_{I}(x_{2})+A_{\nu }(x_{2})J^{\nu }(x_{2}))[dx]%
\end{array}
\label{e29}
\end{equation}%
in the Bogoliubov book\cite{BOG1975} notation, which becomes

\begin{equation}
\begin{array}{l}
A_{1}^{\mu }(x)\simeq -e\tint D_{0}^{c}(x-y)\overset{\ast }{\Phi }%
_{1p^{\prime }}:\overline{\psi _{e}}(y)\gamma ^{\mu }\psi _{e}(y):\Phi
_{1p}d^{4}y \\ 
\\ 
\Phi _{1p}=(2\pi )^{\frac{3}{2}}\overset{\ast }{a_{\nu }^{+}}(%
\overrightarrow{p})\Phi _{0}%
\end{array}
\label{e30}
\end{equation}

But (\ref{e20}) is singular at the ring with radius $a$, while the above
perturbative term is singular at the point $\overrightarrow{x}=0$, which
emerge after an expansion of (\ref{e20}) and the definition of $r$ in powers
of $a=\frac{\hbar }{2m}$. The emergence of the Plank constant $\hbar $
strongly indicates\ that the electromagnetic field (\ref{e20}) includes the
contributions of loop diagrams. This observation makes clear that if PCFT is
the background dynamics of the Bogoliubov harmonic expansion, there must be
a shelf-consistency condition \ between the "classical" solutions (like the
Kerr-Newman manifold) and the summation of the "quantum" series. Besides,
the Bogoliubov free-field expansion should be the positive energy
representations of the conformal group\cite{Mack} with finite component
fields\cite{MackSalam}.

\section{ELECTROWEAK GAUGE\ FIELD}

\setcounter{equation}{0}

We saw that the symplectic 2-form (\ref{v18}) is defined in the unbounded $%
SU(2,2)$ symmetric Siegel domain $X^{\dag }E_{U}X>0$. Its boundary is $%
\mathbb{R}
^{4}$ with a Minkowski induced metric. The corresponding bounded realization 
$Y^{\dag }E_{B}Y>0$ has is $U(2)$. The "natural U(2)" LCR-structure is \ 
\begin{equation}
\begin{array}{l}
e=-iw^{-1}dw=:%
\begin{pmatrix}
\ell & \overline{m} \\ 
m & n%
\end{pmatrix}%
\quad ,\quad de-ie\wedge e=0 \\ 
\\ 
d\ell =im\wedge \overline{m}\quad ,\quad dn=-im\wedge \overline{m}\quad
,\quad dm=i(\ell -n)\wedge m%
\end{array}
\label{n1}
\end{equation}%
This form strongly suggests to osculate the LCR-structure with the $U(2)$\
group. The first step of that is to cast a LCR-tetrad into the hermitian
matrix \ 
\begin{equation}
\begin{array}{l}
e^{\prime }:=%
\begin{pmatrix}
\ell ^{\prime } & \overline{m^{\prime }} \\ 
m^{\prime } & n^{\prime }%
\end{pmatrix}%
=i(\partial -\overline{\partial })%
\begin{pmatrix}
\rho _{11} & \rho _{12} \\ 
\overline{\rho _{12}} & \rho _{22}%
\end{pmatrix}
\\ 
\end{array}
\label{n1a}
\end{equation}%
where $\rho _{ij}$\ is the hermitian LCR embedding relations (\ref{i6}).
Following the Maurer-Cartan procedure we consider the hermitian matrix $%
e^{\prime }$ an element of $u(2)$ Lie algebra, i.e. a $U(2)$ connection with
non-vanishing curvature. The connection and the corresponding curvature are
\ 
\begin{equation}
\begin{array}{l}
B=B_{I\mu }dx^{\mu }t_{I}=%
\begin{pmatrix}
\ell ^{\prime } & \overline{m^{\prime }} \\ 
m^{\prime } & n^{\prime }%
\end{pmatrix}%
\quad ,\quad \lbrack t_{I},t_{J}]=iC_{IJK}t_{K} \\ 
\\ 
F=dB-iB\wedge B\ \longrightarrow DF:=\ dF+iB\wedge F-iF\wedge B=0 \\ 
\end{array}
\label{n1b}
\end{equation}%
where $t_{J}$ are generators of $U(2)$. Apparently a gauge transformation
breaks the tetrad-Weyl symmetry. That is, the $U(2)$ transformation is
expected to transform LCR-structures to other LCR-structures, like the weak $%
U(2)$ transforms electron to its neutrino and vice-versa. The $SO(1,3)$
transformation of the null tetrads also breaks the LCR-structure. Therefore
we chose the LCR-tetrad $e^{\prime }$ to be such that $\Phi _{1}^{\prime
}=1=-\Phi _{2}^{\prime }$.\ That is, we partly fix the tetrad-Weyl symmetry
for non-trivial LCR-structures with $\Phi _{1}\neq 0\neq \Phi _{2}$. Recall
the general tetrad-Weyl transformation is \ 
\begin{equation}
\begin{array}{l}
\ell ^{\prime }=\Lambda \ell \quad ,\quad n^{\prime }=Nn\quad ,\quad
m^{\prime }=Mm \\ 
\\ 
Z_{1}^{\prime }=Z_{1}+d(\ln \Lambda )\quad ,\quad \Phi _{1}^{\prime }=\frac{%
\Lambda }{M\overline{M}}\Phi _{1} \\ 
Z_{2}^{\prime }=Z_{2}+d(\ln N)\quad ,\quad \Phi _{2}^{\prime }=\frac{N}{M%
\overline{M}}\Phi _{2} \\ 
Z_{3}^{\prime }=Z_{3}+d(\ln M)\quad ,\quad \Phi _{3}^{\prime }=\frac{M}{%
\Lambda N}\Phi _{3}%
\end{array}
\label{n1c}
\end{equation}

In the case of the following generators \ 
\begin{equation}
\begin{array}{l}
t_{0}=I\quad ,\quad t_{k}=\frac{\sigma _{k}}{2}\ \rightarrow \
C_{ijk}=\epsilon _{ijk} \\ 
\end{array}
\label{n1d}
\end{equation}%
we have\ 
\begin{equation}
\begin{array}{l}
B_{0\mu }+\frac{1}{2}B_{3\mu }=\ell _{\mu }^{\prime }\quad ,\quad B_{0\mu }-%
\frac{1}{2}B_{3\mu }=n_{\mu }^{\prime }\quad ,\quad \frac{1}{2}(B_{1\mu
}+iB_{2\mu })=m_{\mu }^{\prime } \\ 
\\ 
F_{0\mu \nu }=\partial _{\mu }B_{0\nu }-\partial _{\nu }B_{0\mu } \\ 
F_{i\mu \nu }=\partial _{\mu }B_{i\nu }-\partial _{\nu }B_{i\mu }-\epsilon
_{ijk}B_{j\mu }B_{k\nu }%
\end{array}
\label{n1e}
\end{equation}%
The standard model relations between the $U(1)$ gauge potential $B_{0\mu }$
and the $SU(2)$ gauge potentials $B_{j\mu }$ suggest us to identify the
electromagnetic potential $A_{\mu }$ with $\ell _{\mu }^{\prime }$, the
neutral potential $Z_{\mu }$ with $n_{\mu }^{\prime }$ and the charged
potential $W_{\mu }$ with $m_{\mu }^{\prime }$. Besides, the relative
invariants are apparently related to the Higgs field.

In the case of the electron LCR-tetrad (\ref{e4}) the three relative
invariants are (\ref{e25}) 
\begin{equation}
\begin{array}{l}
\Phi _{1}=\frac{-2a\cos \theta }{r^{2}+a^{2}\cos ^{2}\theta } \\ 
\Phi _{2}=-\frac{(r^{2}+a^{2}+h)a\cos \theta }{(r^{2}+a^{2}\cos ^{2}\theta
)^{2}} \\ 
\Phi _{3}=\frac{2iar\sin \theta }{\sqrt{2}(r+ia\cos \theta )^{2}(r-ia\cos
\theta )} \\ 
\end{array}
\label{n1f}
\end{equation}%
We first make the tetrad-Weyl transformation to reach the conditions $\Phi
_{1}^{\prime }=1=-\Phi _{2}^{\prime }$. We find \ \ 
\begin{equation}
\begin{array}{l}
N=-\frac{r^{2}+a^{2}\cos ^{2}\theta }{r^{2}+a^{2}+h}\Lambda \\ 
M\overline{M}=-\frac{2a\cos \theta }{r^{2}+a^{2}\cos ^{2}\theta }\Lambda \\ 
\end{array}
\label{n1g}
\end{equation}%
The electromagnetic dressing (\ref{e4a}) is found with $\Lambda =\frac{qr}{%
r^{2}+a^{2}\cos ^{2}\theta }$. Then the electroweak connection $B$ (\ref{n1e}%
) is found with\ \ 
\begin{equation}
\begin{array}{l}
\Lambda =\frac{qr}{r^{2}+a^{2}\cos ^{2}\theta } \\ 
N=-\frac{qr}{4\pi (r^{2}+a^{2}+h)} \\ 
M\overline{M}=-\frac{qra\cos \theta }{2\pi (r^{2}+a^{2}\cos ^{2}\theta )^{2}}
\\ 
\end{array}
\label{n2}
\end{equation}%
up to an $M$ phase tetrad-Weyl transformation. That is, we find the
following electroweak potentials (dressings) of the electron%
\begin{equation}
\begin{array}{l}
A=\frac{qr}{4\pi (r^{2}+a^{2}\cos ^{2}\theta )}(dt-dr-a\sin ^{2}\theta
d\varphi ) \\ 
Z=\frac{-qr}{8\pi (r^{2}+a^{2}\cos ^{2}\theta )}(dt+\frac{r^{2}+2a^{2}\cos
^{2}\theta -a^{2}-h}{r^{2}+a^{2}+h}dr-a\sin ^{2}\theta d\varphi ) \\ 
W=\frac{-M}{\sqrt{2}(r+ia\cos \theta )}[-ia\sin \theta \
(dt-dr)+(r^{2}+a^{2}\cos ^{2}\theta )d\theta + \\ 
\qquad \qquad +i\sin \theta (r^{2}+a^{2})d\varphi ]%
\end{array}
\label{n3}
\end{equation}%
where the tetrad-Weyl factor $M$ will be computed below through the Higgs
dressing.

The third (complex) relative invariant $\Phi _{3}^{\prime }$ (\ref{n1c}) is
not completely fixed. Its phase is absorbed by $W$ and the remaining scalar
real field $\Phi _{3}^{\prime }$ will be finally related with the electron
Higgs potential\ \ 
\begin{equation}
\begin{array}{l}
M=\frac{i}{r-ia\cos \theta }[\frac{qra\cos \theta }{2\pi (r^{2}+a^{2}\cos
^{2}\theta )^{2}}]^{\frac{1}{2}} \\ 
\Phi _{3}^{\prime }=\frac{2\sin \theta (r^{2}+a^{2}+h)}{r^{2}+a^{2}\cos
^{2}\theta }[\frac{\pi a^{3}\cos \theta }{q^{3}r}]^{\frac{1}{2}} \\ 
\end{array}
\label{n4}
\end{equation}

We see that Einstein's $SO(1,3)$ connection and the $U(2)$ gauge field are
directly related to the LCR-tetrad and the Higgs field is related with the $%
\Phi _{i}$\ relative invariants of the LCR-structure. Hence all the known
fields of the electroweak interactions exist in the LCR-structure. They are
introduced in the algorithm of Bogoliubov, Epstein-Glaser, Scharf and
collaborators as asymptotic free fields. The recursive relations provide\cite%
{Sch2} all the relations between the masses and the coupling constants.

The idea of Scharf and collaborators was to exploit the fact that the
S-matrix is an operator valued distribution which is expanded to the
operator valued distributions of precise free fields (Poincar\'{e}
representations) in the rigged Hilbert-Fock space of tembered distributions.
In this framework we have to consider the operator form of the gauge
transformations\ \ 
\begin{equation}
\begin{array}{l}
A^{\prime \mu }=e^{-i\lambda Q}A^{\mu }e^{-i\lambda Q} \\ 
\multicolumn{1}{c}{\Downarrow} \\ 
\delta A^{\mu }=-i\lambda \lbrack Q,A^{\mu }]%
\end{array}
\label{n5}
\end{equation}%
where $Q$ is a nilpotent operator. For every gauge field $A^{\mu }$ we
introduce the anticommuting ghost fields $u(x)$, $\widetilde{u}(x)$ and the
implied gauge field triplet acquire the transformations\ \ 
\begin{equation}
\begin{array}{l}
d_{Q}A^{\mu }=[Q,A^{\mu }]=i\partial ^{\mu }u \\ 
d_{Q}u=\{Q,u\}=0 \\ 
d_{Q}\widetilde{u}=\{Q,\widetilde{u}\}=-i\partial _{\mu }A^{\mu }%
\end{array}
\label{n6}
\end{equation}%
The invariance of the S-matrix essentially relative to this deformation
essentially protect the generation of non-physical modes. That is, the
physical Fock space is the $\ker (Q^{\dag }Q+QQ^{\dag })$.

Applying this algorithm to our case of a pair of asymptotic (free) massive
fermions $\psi _{e}$, $\psi _{\nu }$, and a $U(2)$ gauge field with one
massless and three massive, we start from general interaction lagrangian.
The leptonic and gauge field variations are\ \ 
\begin{equation}
\begin{array}{l}
d_{Q}\psi _{e}=0\ ,\quad d_{Q}\psi _{\nu }=0 \\ 
d_{Q}A^{\mu }=i\partial ^{\mu }u_{0}\ ,\quad d_{Q}W_{1,2}^{\mu }=i\partial
^{\mu }u_{1,2}\ ,\quad d_{Q}Z^{\mu }=i\partial ^{\mu }u_{3} \\ 
\end{array}
\label{n7}
\end{equation}%
The gauge variations for the Higgs $\phi $ and the unphysical scalar fields $%
\Phi _{1,2,3}$ are\ \ 
\begin{equation}
\begin{array}{l}
d_{Q}\phi =0\ ,\quad d_{Q}\Phi _{1,2}=im_{W}u_{1,2}\ ,\quad d_{Q}\Phi
_{3}=im_{Z}u_{3} \\ 
\end{array}
\label{n8}
\end{equation}%
The gauge variations for the fermionic ghosts are\ \ 
\begin{equation}
\begin{array}{l}
d_{Q}u_{b}=0\ ,\quad b=0,1,2,3 \\ 
d_{Q}\widetilde{u}_{0}=-i\partial _{\mu }A^{\mu }\ ,\quad d_{Q}\widetilde{u}%
_{1,2}=-i\partial _{\mu }(W_{1,2}^{\mu }+m_{W}\Phi _{1,2}) \\ 
d_{Q}\widetilde{u}_{1,2}=-i\partial _{\mu }(Z^{\mu }+m_{Z}\Phi _{3})%
\end{array}
\label{n9}
\end{equation}%
After very long calculations Scharf and collaborators\cite{Sch2} succeeded
to reproduce the standard model lagrangian with the proper relations between
coupling constants and masses! Besides the application of this algorithm to
the gravitational interactions, they essentially found that the protection
from the unphysical modes restricts the action to the Einstein-Hilbert form,
because simply it is the only scalar form without second order derivatives,
which have negative norm states.

\subsection{On the origin of the leptonic generations}

The electron and its neutrino LCR-structures derived from the linear
trajectory $\xi ^{a}=v^{a}\tau +c^{a}$ or equivalently the quadratic Kerr
polynomial \ 
\begin{equation}
\begin{array}{l}
K(Z)=iZ^{0}Z^{0}[(v^{0}-v^{3})(c^{1}+ic^{2})-(v^{1}+iv^{2})(c^{0}-c^{3})]+
\\ 
\qquad
+iZ^{0}Z^{1}[(v^{0}+v^{3})(c^{0}-c^{3})-(v^{1}-iv^{2})(c^{1}+ic^{2})]-Z^{0}Z^{2}(v^{1}+iv^{2})-
\\ 
\qquad
-Z^{0}Z^{3}(v^{0}-v^{3})+iZ^{1}Z^{1}[(v^{1}-iv^{2})(c^{0}-c^{3})-(v^{0}+v^{3})(c^{1}-ic^{2})]+
\\ 
\qquad +Z^{1}Z^{2}(v^{0}+v^{3})+Z^{1}Z^{3}(v^{1}-iv^{2}) \\ 
\end{array}
\label{n10}
\end{equation}%
with $c^{a}$ generally complex. This is the most general quadratic Kerr
polynomial which incorporates all the parameters of the Poincar\'{e}
representation. The singular points of this quadratic surface satisfy the
relations \ 
\begin{equation}
\begin{array}{l}
\partial _{n}K(Z)=0\quad ,\quad Z^{n}\neq 0 \\ 
\end{array}
\label{n11}
\end{equation}%
We finally find that there are the following two cases \ 
\begin{equation}
\begin{array}{l}
1st:\quad If\ \ v^{a}v^{b}\eta _{ab}\neq 0\quad the\ surface\ is\ irreducible
\\ 
\\ 
2nd:\quad If\ \ v^{a}v^{b}\eta _{ab}=0\quad the\ surface\ is\ reducible \\ 
\end{array}
\label{n12}
\end{equation}%
The first case gives the electron and positron LCR-solitons and the second
reducible surface gives the left-handed chiral part of the neutrino. The
electron LCR-structure is determined with an irreducible quadratic
polynomial and the neutrino LCR-structure is determined with the
corresponding reducible quadratic polynomial. Their gravitational metrics $%
g_{\mu \nu }$\ admit only two geodetic and shear free null congruences and
are type-D spacetimes in the Petrov classification. They constitute the
electronic generation of the observed standard model. In order to look for
the other two generations (the muon and tau families) we have to describe
the general framework of the solutions of the pseudo-conformal manifolds and
their topological obstructions.

The grassmannian manifold $G_{4,2}$ is the set of lines of $CP^{3}$. A point
of $G_{4,2}$ is a line of $CP^{3}$, which is determined by the two columns $%
X^{n1}$ and $X^{n2}$ viewed as two points of $CP^{3}$. The linear
transformation $SL(4,C)$, which applies from the left side, preserves the
LCR-structure. The $SL(2,C)$ linear transformation, which applies from the
right side, preserves the line of $CP^{3}$, but it does not preserve the
LCR-structure. Hence, in the present formalism the points of the Minkowski
spacetime are lines of $CP^{3}$.

Every two intersection points of the line and the surface $K(Z^{m})=0$
determine a LCR-structure. In every affine space of $G_{4,2}$, the line is
projectively represented with a $2\times 2$ matrix $r=r_{a}\sigma
_{A^{\prime }A}^{a}$ , which, after the application of the implicit function
theorem for the solution of the four (real) relations (\ref{v5}) of the
LCR-structure, takes the form \ 
\begin{equation}
\begin{array}{l}
r^{a}=x^{a}+iy^{a}(x^{b}) \\ 
\end{array}
\label{n13}
\end{equation}%
The LCR-structure with vanishing $y^{a}(x^{b})=0$ are compatible with the
Minkowski metric. Therefore we may consider this imaginary part as the
gravitational content of the LCR-structure. Hence, the linearized Einstein
gravity approximation projects the LCR-structure down to its "flatprint".
The well known to general relativists $SO(1,3)$ local transformations of the
null tetrad, which preserve the metric, coincide with the line preserving $%
SL(2,C)$ transformation. In fact the quartic polynomial (\ref{g7}) is the
maximum degree Kerr polynomial permitted by a regular Einstein metric, which
admits geodetic and shear free null congruences $\ell ^{\mu }\partial _{\mu
} $\ and $n^{\mu }\partial _{\mu }$. Taking into consideration that the
degree of the Kerr polynomial is a topological invariant of the
corresponding surface of $CP^{3}$, we expect the existence of two more
chiral currents to be permitted in addition to the above studied quadratic
ones. Those determined by cubic Kerr polynomials, which we identify with the
muon generation, and those determined by quartic Kerr polynomials, which we
identify with the tau generation.

The preceding analysis indicates to correlate the lepton numbers with the
degrees of the Kerr polynomial. Then the limitation of the number of
generations is imposed by the Einstein gravity. At each spacetime point,
which is determined by a line of $CP^{3}$, there are at most four
intersections between the Kerr surface and the line. The 1st generation ($%
e,\nu _{e}$) corresponds to quadratic surfaces, which we have extensively
studied before.

Notice that the Kerr polynomial of the curved LCR-manifold is the same with
its corresponding flatprint. Therefore we will make the computation of the
Hopf in this simplified case. Every column $i$ of the $G_{4,2}$ homogeneous
coordinates $X^{ni}$ determine a complex function $\lambda ^{Ai}(x)$ in $%
S^{2}$. That is, for any LCR-structure we have two functions 
\begin{equation}
\begin{array}{l}
\lambda ^{Ai}(x):S^{1}\times S^{3}\rightarrow S^{2} \\ 
\end{array}
\label{n14}
\end{equation}%
understood to be in the bounded $U(2)$ realization of spacetime. It is known
that the homotopy group $\pi _{1}(S^{2})$ is trivial but $\pi _{3}(S^{2})=%
\mathbb{Z}
$. The Hopf invariant is determined using the sphere volume 2-form 
\begin{equation}
\begin{array}{l}
\omega =\frac{i}{2\pi }\frac{d\lambda \wedge d\overline{\lambda }}{%
(1+\lambda \overline{\lambda })^{2}} \\ 
\end{array}
\label{n15}
\end{equation}%
which is closed. This implies that in $S^{3}$ there is an exact 1-form $%
\omega _{1}$ such that $\omega =d\omega _{1}$. Then the Hopf invariant of $%
\lambda (x)$ is 
\begin{equation}
\begin{array}{l}
H(\lambda )=\int \lambda ^{\ast }(\omega )\wedge \omega _{1} \\ 
\end{array}
\label{n16}
\end{equation}%
The Hopf invariants (linking numbers) of the electron and neutrino
LCR-structures have been computed\cite{RAG2008b} and found to be $\pm \frac{a%
}{|a|}$. The higher Hopf invariants may be viewed as a composition of the
above simple ones and the higher $S^{2}\rightarrow S^{2}$.

In the flatprint electron LCR-structure, the relation between the cartesian
coordinates and the structure coordinates are%
\begin{equation}
\begin{array}{l}
x^{0}=t \\ 
x^{1}+ix^{2}=(r-ia)\sin \theta e^{i\varphi } \\ 
x^{3}=r\cos \theta \\ 
\end{array}
\label{n17}
\end{equation}%
For constant time, the (left) causal ray $\ell ^{\mu }(r)$ is%
\begin{equation}
\begin{array}{l}
x^{0}=t=0 \\ 
x^{1}=(r\cos \varphi +a\sin \varphi )\sin \theta \\ 
x^{2}=(r\sin \varphi -a\cos \varphi )\sin \theta \\ 
x^{3}=r\cos \theta \\ 
\end{array}
\label{n18}
\end{equation}%
Recall that the entire $SU(2)$\ space is covered by considering $r\in
(-\infty ,+\infty )$. This 3-dimensional space may be considered as the
initial data of the electron LCR-manifold.\ 

Let us now consider the following initial data%
\begin{equation}
\begin{array}{l}
x^{0}=t=0 \\ 
x^{1}=(r\cos k\varphi +a\sin k\varphi )\sin \theta \\ 
x^{2}=(r\sin k\varphi -a\cos k\varphi )\sin \theta \\ 
x^{3}=r\cos \theta \\ 
\\ 
r\in (-\infty ,+\infty ),\quad \theta \in (0,\pi ),\quad \varphi \in (0,2\pi
),\quad k\in 
\mathbb{Z}%
\end{array}
\label{n19}
\end{equation}%
and compute the linking number of the circles%
\begin{equation}
\begin{array}{l}
\overrightarrow{x^{\prime }}=(0,\ 0,\ r)\quad ,\quad \ \theta =0,\ \varphi =0
\\ 
\overrightarrow{x}=(a\sin k\varphi ,\ -a\cos k\varphi ,\ 0)\quad ,\quad \
r=0,\ \theta =\frac{\pi }{2} \\ 
\\ 
l=\frac{1}{4\pi }\tiint \frac{\epsilon _{ijk}(x^{\prime
i}-x^{i})dx^{j}\wedge dx^{\prime k}}{[\tsum\limits_{i}(x^{\prime
i}-x^{i})^{2}]^{\frac{3}{2}}}%
\end{array}
\label{n20}
\end{equation}%
\begin{equation}
\begin{array}{l}
\overrightarrow{x^{\prime }}=(0,\ 0,\ r)\quad ,\quad \ \theta =0,\ \varphi =0
\\ 
\overrightarrow{x}=(a\sin k\varphi ,\ -a\cos k\varphi ,\ 0)\quad ,\quad \
r=0,\ \theta =\frac{\pi }{2} \\ 
\\ 
l=\frac{ka^{2}}{4\pi }\tiint \frac{drd\varphi }{(a^{2}+r^{2})^{\frac{3}{2}}}=%
\frac{ka}{2|a|}\tint\limits_{-\infty }^{\infty }\frac{dr^{\prime }}{%
(1+r^{\prime 2})^{\frac{3}{2}}}=\frac{ka}{|a|}%
\end{array}
\label{n21}
\end{equation}%
It apparently counts how many times the circle of the ring singularity winds
around $\overrightarrow{x^{\prime }}(r)$, the $\rho $-closed curve of $SU(2)$%
.

This theoretical reasoning, based on $\pi _{2}(S^{3})=%
\mathbb{Z}
$, needs to be accompanied with the reason why the observed leptonic
generations are three and not infinite as it is suggested. The above
computations are done in zero gravity limit. Hence the restriction of the
number of generations should be caused by gravity through the following
reasoning.

The gravitational dressing of the elementary particles satisfies Einstein's
equations through the metric $g_{\mu \nu }$, which admits geodetic and
shear-free null congruences. These congruences are principal null directions
of the Weyl tensor which is formally written%
\begin{equation}
\begin{array}{l}
\ell ^{\mu }=\overline{\kappa }^{A^{\prime }}\sigma _{A^{\prime
}A}^{b}\kappa ^{A}e_{b}^{\cdot \mu }\quad ,\quad \ \Psi _{ABCD}\kappa
^{A}\kappa ^{B}\kappa ^{C}\kappa ^{D}=0 \\ 
\end{array}
\label{n22}
\end{equation}%
in the Newman-Penrose formalism. Hence the number of gravitational principal
directions cannot exceed four and subsequently the degree of the Kerr
polynomial cannot be higher than four.

Let us now pose the question "how many static massive and (stationary)
massless representations of polynomial multiplets of a given degree exist?".
The knowledge of the Poincar\'{e} group permit us to answer this question by
simply noticing that such a multiplet must contain a polynomial, which
admits the infinitesimal z-rotations and time-translation as automorphisms.

I have worked it out\cite{RAG2008b} and found that only the irreducible
quadratic surface (\ref{g10a}) determines a polynomial multiplet. If our
hypothesis that the lepton numbers are the topological degrees of the Kerr
polynomial, then we have to conclude that the muon and tau are unstable
solitonic surfaces, which is also observed. Therefore it would be very
interesting to find solitonic third and fourth degree surfaces.

\section{HADRONIC\ SECTOR\ AND\ CONFINEMENT}

\setcounter{equation}{0}

The following tetrad-Weyl invariant gauge field equations 
\begin{equation}
\begin{array}{l}
\frac{1}{\sqrt{-g}}(D_{\mu })_{ij}[\sqrt{-g}(\Gamma ^{\mu \nu \rho \sigma
}\mp \overline{\Gamma ^{\mu \nu \rho \sigma }})F_{j\rho \sigma }]=0 \\ 
\\ 
\Gamma ^{\mu \nu \rho \sigma }=\frac{1}{2}[(\ell ^{\mu }m^{\nu }-\ell ^{\nu
}m^{\mu })(n^{\rho }\overline{m}^{\sigma }-n^{\sigma }\overline{m}^{\rho
})+(n^{\mu }\overline{m}^{\nu }-n^{\nu }\overline{m}^{\mu })(\ell ^{\rho
}m^{\sigma }-\ell ^{\sigma }m^{\rho })] \\ 
(D_{\mu })_{ij}=\delta _{ij}\partial _{\mu }-\gamma f_{ikj}A_{k\mu }%
\end{array}
\label{h1}
\end{equation}%
may be written down. The do not have sources. Therefore I find convenient to
give them PDEs the following forms 
\begin{equation}
\begin{array}{l}
(-)\quad \rightarrow \quad \frac{1}{\sqrt{-g}}(D_{\mu })_{ij}\{\sqrt{-g}%
[(\ell ^{\mu }m^{\nu }-\ell ^{\nu }m^{\mu })(n^{\rho }\overline{m}^{\sigma
}F_{j\rho \sigma })+ \\ 
\qquad \qquad +(n^{\mu }\overline{m}^{\nu }-n^{\nu }\overline{m}^{\mu
})(\ell ^{\rho }m^{\sigma }F_{j\rho \sigma })]\}=-k_{i}^{\nu } \\ 
\\ 
(+)\quad \rightarrow \quad \frac{1}{\sqrt{-g}}(D_{\mu })_{ij}\{\sqrt{-g}%
[(\ell ^{\mu }m^{\nu }-\ell ^{\nu }m^{\mu })(n^{\rho }\overline{m}^{\sigma
}F_{j\rho \sigma })+ \\ 
\qquad \qquad +(n^{\mu }\overline{m}^{\nu }-n^{\nu }\overline{m}^{\mu
})(\ell ^{\rho }m^{\sigma }F_{j\rho \sigma })]\}=-ik_{i}^{\nu }%
\end{array}
\label{h2}
\end{equation}%
where $k_{i}^{\nu }(x)$ is a real vector field which we will consider as
distributional (localized) sources. That is, taking into account that the
sum of the two terms is self-dual 
\begin{equation}
\begin{array}{l}
\Gamma ^{\mu \nu \rho \sigma }F_{j\rho \sigma }=:G_{j}^{\mu \nu }-i\ast
G_{j}^{\mu \nu } \\ 
\end{array}
\label{h3}
\end{equation}%
we may find a real gauge field with electric sources in the first case (-),
and magnetic sources in the second case (+). The framework of distributional
solitons seems to be studied the problem of hadronic sector in PCFT.

We will here study the properties of such non-null abelian solitons found%
\cite{RAG2018b} in the case of the flatprint static LCR-structure (\ref{e3}%
). It is more convenient to make calculations using the differential forms
of (\ref{h2})

\begin{equation}
\begin{array}{l}
d\{(n^{\rho }\overline{m}^{\sigma }F_{\rho \sigma })\ell \wedge m+(\ell
^{\rho }m^{\sigma }F_{\rho \sigma })n\wedge \overline{m}\}=i\ast k \\ 
\end{array}
\label{h4}
\end{equation}%
Using the Stoke's theorem procedure\cite{RAG2018b}, we find the
non-vanishing closed 2-forms (with real sources) in the case of flatprint
massive static LCR-tetrad

\begin{equation}
\begin{array}{l}
(L^{\rho }M^{\sigma }F_{\rho \sigma })=\frac{C^{\prime \prime }(r+ia\cos
\theta )}{(r^{2}+a^{2})\sin \theta }\quad ,\quad (N^{\rho }\overline{M}%
^{\sigma }F_{\rho \sigma })=\frac{C^{\prime }}{\sin \theta (r-ia\cos \theta )%
} \\ 
F_{\rho \sigma }=\partial _{\rho }A_{\sigma }-\partial _{\sigma }A_{\rho }%
\end{array}
\label{h5}
\end{equation}%
where $C^{\prime }$ and $C^{\prime \prime }$ are arbitrary complex
constants, which are fixed using Stokes' theorem and the reality conditions
for gluonic sources. We also assume that 
\begin{equation}
\begin{array}{l}
\lbrack (L^{\mu }N^{\nu }-M^{\mu }\overline{M}^{\nu })F_{j\mu \nu }]=0 \\ 
\end{array}
\label{h6}
\end{equation}%
because it cannot be determined by the sources. That is

\begin{equation}
\begin{array}{l}
F_{\rho \sigma }=-\frac{C^{\prime }}{\sin \theta (r-ia\cos \theta )}(L_{\rho
}M_{\sigma }-L_{\sigma }M_{\sigma })-\frac{C^{\prime \prime }(r+ia\cos
\theta )}{(r^{2}+a^{2})\sin \theta }(N_{\rho }\overline{M}_{\sigma
}-N_{\sigma }\overline{M}_{\sigma })+c.c. \\ 
\end{array}
\label{h7}
\end{equation}

For the static flatprint LCR-tetrad the solutions have the explicit forms

\begin{equation}
\begin{array}{l}
F-i\ast F:=-\frac{2C^{\prime }}{\sin \theta (r-ia\cos \theta )}L\wedge M-%
\frac{2C^{\prime \prime }(r+ia\cos \theta )}{(r^{2}+a^{2})\sin \theta }%
N\wedge \overline{M}= \\ 
\quad =\frac{2C^{\prime }+C^{\prime \prime }}{\sqrt{2}}[\frac{-ia}{%
r^{2}+a^{2}}dt\wedge dr+\frac{1}{\sin \theta }dt\wedge d\theta +\frac{%
a^{2}\sin \theta }{r^{2}+a^{2}}dr\wedge d\theta \\ 
\quad -idr\wedge d\varphi +a\sin \theta d\theta \wedge d\varphi ]+ \\ 
\quad +\frac{2C^{\prime }-C^{\prime \prime }}{\sqrt{2}}[\frac{ia}{%
(r^{2}+a^{2})}dt\wedge dr+idt\wedge d\varphi -\frac{r^{2}+a^{2}\cos
^{2}\theta }{(r^{2}+a^{2})\sin \theta }dr\wedge d\theta ]%
\end{array}
\label{h8}
\end{equation}%
After a straightforward calculation I find

\begin{equation}
\begin{array}{l}
\tint\limits_{t,r=const}[\frac{-2C^{\prime }}{\sin \theta (r-ia\cos \theta )}%
L\wedge M+\frac{-2C^{\prime \prime }(r+ia\cos \theta )}{(r^{2}+a^{2})\sin
\theta }N\wedge \overline{M}]=\frac{(2C^{\prime }+C^{\prime \prime })4\pi a}{%
\sqrt{2}}=:-i\gamma \\ 
\end{array}
\label{h9}
\end{equation}%
where $\gamma $ is the real hadronic constant. Assuming $2C^{\prime
}-C^{\prime \prime }=0$ (because it is not determined by the gluonic
charge), the arbitrary constants are completely fixed and the solutions are

\begin{equation}
\begin{array}{l}
F=\frac{-\gamma }{4\pi a}[\frac{a}{r^{2}+a^{2}}dt\wedge dr+dr\wedge d\varphi
]= \\ 
\qquad =d[\frac{-\gamma }{4\pi a}(\tan ^{-1}\frac{r}{a}dt+rd\varphi )] \\ 
\ast F=\frac{\gamma }{4\pi }[\frac{1}{a\sin \theta }dt\wedge d\theta +\frac{%
a\sin \theta }{r^{2}+a^{2}}dr\wedge d\theta +\sin \theta d\theta \wedge
d\varphi ]%
\end{array}
\label{h10}
\end{equation}%
where the corresponding gluonic potential (quark dressing) been apparent. In
cartesian coordinates the potential and the gluonic field strength take the
form%
\begin{equation}
\begin{array}{l}
A^{(g)}=\frac{-\gamma }{4\pi a}(\tan ^{-1}\frac{r}{a}dt+rd\varphi )= \\ 
\qquad =\frac{-\gamma }{4\pi a}(\tan ^{-1}\frac{r}{a}dx^{0}-\frac{%
ax^{1}+rx^{2}}{(x^{1})^{2}+(x^{2})^{2}}dx^{1}+\frac{rx^{1}-ax^{2}}{%
(x^{1})^{2}+(x^{2})^{2}}dx^{2}) \\ 
\\ 
F^{(g)}=\frac{-\gamma }{4\pi a}(\frac{a}{r^{2}+a^{2}}dt\wedge dr+dr\wedge
d\varphi )= \\ 
\qquad =\frac{-\gamma }{4\pi (r^{2}+a^{2})}[\frac{rx^{1}-ax^{2}}{r^{2}+a^{2}}%
dx^{0}\wedge dx^{1}+\frac{ax^{1}+rx^{2}}{r^{2}+a^{2}}dx^{0}\wedge dx^{2}+%
\frac{x^{3}}{r}dx^{0}\wedge dx^{3}]+ \\ 
\qquad +\frac{-\gamma }{4\pi ar}[dx^{1}\wedge dx^{2}+\frac{%
x^{3}(ax^{1}+rx^{2})}{r((x^{1})^{2}+(x^{2})^{2})}dx^{1}\wedge dx^{3}-\frac{%
x^{3}(rx^{1}-ax^{2})}{r((x^{1})^{2}+(x^{2})^{2})}dx^{2}\wedge dx^{3}]%
\end{array}
\label{h11}
\end{equation}%
while the form of the electric potential is%
\begin{equation}
\begin{array}{l}
A^{(e)}=\frac{qr^{3}}{4\pi (r^{4}+a^{2}(x^{3})^{2})}(dx^{0}-\frac{%
rx^{1}-ax^{2}}{r^{2}+a^{2}}dx^{1}-\frac{rx^{2}+ax^{1}}{r^{2}+a^{2}}dx^{2}-%
\frac{x^{3}}{r}dx^{3}) \\ 
\end{array}
\label{h12}
\end{equation}%
It has a line singularity along the z-axis like the Dirac magnetic monopole,
but its asymptotic behavior is different. The "electric" and "magnetic"
parts are

\begin{equation}
\begin{array}{l}
A_{0}^{(g)}=\frac{-\gamma _{j}}{4\pi a}(\tan ^{-1}\frac{r}{a}) \\ 
\overrightarrow{A}^{(g)}=\frac{-\gamma }{4\pi }[\frac{r}{a}\frac{%
-x^{2}dx^{1}+x^{1}dx^{2}}{(x^{1})^{2}+(x^{2})^{2}}-\frac{1}{2}d\ln
((x^{1})^{2}+(x^{2})^{2})] \\ 
\\ 
r=\pm \left\{ \frac{(x^{1})^{2}+(x^{2})^{2}+(x^{3})^{2}-a^{2}}{2}+\sqrt{[%
\frac{(x^{1})^{2}+(x^{2})^{2}+(x^{3})^{2}-a^{2}}{2}]^{2}+a^{2}(x^{3})^{2}}%
\right\} ^{\frac{1}{2}}%
\end{array}
\label{h13}
\end{equation}%
where the last term of $\overrightarrow{A}^{(g)}$ is a singular gauge. The
"magnetic" part of the potential is linear indicating confinement.

But the essential difference of the gluonic dressing is its singularity
relative to the angular variable $a$. This does not permit us to make a
harmonic expansion into free fields. The Bogoliubov causal perturbative
approach and its Scharf version cannot been applied to "derive" quantum
effects.

\section{BEYOND\ THE STANDARD\ MODEL}

\setcounter{equation}{0}

Standard model may be considered as the great success of modern fundamental
physics. In the beginning we thought that extending the $U(2)$ electroweak
group to larger internal groups, like $SU(5)$, we could incorporate the
hadronic sector. The experimental disagreement with the decay rate of the
proton blocked these expectations. The next attempt was the consideration of
supersymmetric groups, which combine bosons and fermions. The recent (last)
experimental result at CERN, that supersymmetric particles do not exist,
blocked this theoretical investigation too. String theory was the last
theoretical framework of the beyond the standard attempts to incorporate
gravity, but its supersymmetric basis was fatal for this theory too.

Pseudo-conformal field theory (PCFT) is purely geometric. Its fundamental
physical principle is the existence of a lorentzian Cauchy-Riemann (LCR)
structure defined in the tangent space (bundle) of a manifold (spacetime)
like the riemannian metric. The existence of (Schwartz) distributional
solutions "needs" the mathematical extension to the tempered distributions
and the harmonic analysis (series) in the corresponding rigged Hilbert space
(Gelfand triple). Recall that the Pythagora computation of hypotenuse
"needed" the completion of the rational numbers $%
\mathbb{Q}
$ to the real numbers $%
\mathbb{R}
$, computed as infinite summations of rational numbers. In this section I
will try to indicate some directions (suggestions) where the explanation of
some "beyond the standard model" may be hidden.

\subsection{On the dark matter and energy}

The dark energy may be a proper normalization cosmological constant in the
Scharf algorithmic graviton field expansion. But my negative results to find
other static or stationary solitons drives me to look for possible geometric
explanations of cosmological dark matter.

The macroscopic study of the universe should be done through the Kaehler
manifold determined by (\ref{v16}). "Dark matter" may be the effect of the
second fundamental form of the embedding of the universe LCR-submanifold in $%
\mathbb{C}
^{4}$.

In order to visualize the embedding consequences of the universe as a real
submanifold of $%
\mathbb{C}
^{4}$, it is convenient to work in real coordinates. In the Eisenhart book%
\cite{Eisen} notation, the Gauss-Codazzi equations have the form%
\begin{equation}
\begin{array}{l}
R_{ijkl}=\tsum\limits_{\sigma =0}^{8}e_{\sigma }(\Omega _{\sigma |ik}\Omega
_{\sigma |jl}-\Omega _{\sigma |il}\Omega _{\sigma |jk})+\overline{R}_{\alpha
\beta \gamma \delta }y_{,i}^{\alpha }y_{,j}^{\beta }y_{,k}^{\gamma
}y_{,l}^{\delta } \\ 
\\ 
\Omega _{\sigma |ij,k}-\Omega _{\sigma |ik,j}=\tsum\limits_{\tau
=0}^{8}e_{\tau }(\mu _{\tau \sigma |k}\Omega _{\tau |ij}-\mu _{\tau \sigma
|j}\Omega _{\tau |ik})+\overline{R}_{\alpha \beta \gamma \delta
}y_{,i}^{\alpha }y_{,j}^{\gamma }y_{,k}^{\delta }\xi _{\sigma }^{\beta } \\ 
\end{array}
\label{b1}
\end{equation}%
where the universe is $V_{4}$ with induced metric $g_{ij}$, and the
enveloping space is $V_{8}$ with metric $a_{\alpha \beta }$. The latin
indices $i,j,k,...$ take values up to four and the greek indices $\alpha
,\beta ,\gamma ,\mu ,...$ take values up to eight. We see that the induced
curvature $R_{ijkl}$ of the surface depends on the second fundamental form $%
\Omega _{\sigma |ik}$ and the curvature $\overline{R}_{\alpha \beta \gamma
\delta }$ of the ambient Kaehler manifold. These relations of the geometric
tensors are essentially implied by their precise dependence on the four real
embedding functions $\rho _{ij}(\overline{z^{b}},z^{c})$.

One of the most striking effects attributed to "dark matter" comes from the
velocities $v^{j}$ of the stars at different distances from the galactic
center. This mathematical problem is studied in paragraph 48 of the
Eisenhart book\cite{Eisen}, and the following relation (48.7 p.165)%
\begin{equation}
\begin{array}{l}
\frac{e_{a}}{\rho _{a}^{2}}=\frac{e_{g}}{\rho _{g}^{2}}+\frac{e}{R^{2}} \\ 
\end{array}
\label{b2}
\end{equation}
where we have the curve curvatures:[ $\frac{1}{\rho _{a}}$ relative to the
ambient metric $a_{\alpha \beta }$, $\frac{1}{\rho _{g}}$ relative to the
spacetime metric $g_{ij}$, and $\frac{1}{R}$ the normal curvature of
spacetime] and $e_{a}$, $e_{g}$ and $e$ are plus or minus one. Apparently
the deviation of the observed trajectory from the geodetic one indicates
that the observed spacetime is not a totally geodetic submanifold.

\subsection{Harmonic expansion in the bounded domain}

The appearance of $\eta _{\mu \nu }$ in the"vacuum" metric and symplectic
form (\ref{v18}) impose the Poincar\'{e} group harmonic expansion. The used
Bogoliubov causal approach (with the Scharf et. al. improvements) are
heavily based on the unbounded realization of the $SU(2,2)$ classical
domain. But the corresponding bounded realization seems to be more "natural"
because the polar decomposition approach covers the entire spacetime (up to
a point), but its maximal subgroup changes to $S(U(2)\times U(2))$. The
Kaehler function is $\det (I-w^{\dag }w)$ with $w$ a general $2\times 2$
matrix which at the $U(2)$ boundary becomes unitary. The relation between
the bounded and unbounded realizations of the conformal group $SU(2,2)$ and
its relation with the Minkowski space distributions has been analyzed by
Ruehl\cite{Ruehl}.

In the bounded realization the maximal subgroup $S(U(2)\times U(2))$ is 
\begin{equation}
\begin{array}{l}
\begin{pmatrix}
Y_{1}^{\prime } \\ 
Y_{2}^{\prime }%
\end{pmatrix}%
=%
\begin{pmatrix}
U_{1} & 0 \\ 
0 & U_{2}%
\end{pmatrix}%
\left( 
\begin{array}{c}
Y_{1} \\ 
Y_{2}%
\end{array}%
\right) \\ 
\\ 
U_{1}\ ,\ U_{2}\in U(2)\quad ,\quad \det (U_{1}U_{2})=1%
\end{array}
\label{b3}
\end{equation}%
where $U_{1}$\ and $U_{2}$\ are independent $U(2)$ elements. Its Cartan
subalgebra is%
\begin{equation}
\begin{array}{l}
H_{1}=%
\begin{pmatrix}
\frac{\sigma _{3}}{2} & 0 \\ 
0 & 0%
\end{pmatrix}%
\quad ,\quad H_{2}=%
\begin{pmatrix}
0 & 0 \\ 
0 & \frac{\sigma _{3}}{2}%
\end{pmatrix}
\\ 
\\ 
H_{0}=%
\begin{pmatrix}
I & 0 \\ 
0 & -I%
\end{pmatrix}%
\end{array}
\label{b4}
\end{equation}%
The $z$-rotation of the Poincar\'{e} algebra takes the form $%
J_{3}=H_{1}+H_{2}$. in the bounded realization. But the evolution operator
in the bounded representation is\cite{Mack} $H_{0}=$ $\frac{1}{2}%
(P^{0}+K^{0})$, which is essentially the generator of $U(1)$ factor of $%
U_{2}(2)$ (or its universal covering $%
\mathbb{R}
$). It does not coincide with the evolution operator $P^{0}$\ of the
unbounded realization. Hence the static quadratic Kerr polynomial with $%
J_{3} $ and $H_{0}$ automorphisms has the form%
\begin{equation}
\begin{array}{l}
(bounded)\qquad K_{B}=AY^{0}Y^{3}+BY^{1}Y^{2} \\ 
\multicolumn{1}{c}{\Updownarrow} \\ 
(unbounded)\qquad K_{U}=\frac{A+B}{2}(X^{0}X^{1}-X^{2}X^{3})+\frac{A-B}{2}%
(X^{1}X^{2}-X^{0}X^{3})%
\end{array}
\label{b5}
\end{equation}%
where we have used the transformation%
\begin{equation}
\begin{array}{l}
\begin{pmatrix}
Y^{0} \\ 
Y^{1} \\ 
Y^{2} \\ 
Y^{3}%
\end{pmatrix}%
=\frac{1}{\sqrt{2}}\left( 
\begin{array}{cc}
I & I \\ 
I & -I%
\end{array}%
\right) 
\begin{pmatrix}
X^{0} \\ 
X^{1} \\ 
X^{2} \\ 
X^{3}%
\end{pmatrix}
\\ 
\end{array}
\label{b6}
\end{equation}%
between the bounded coordinates $Y^{n}$ and the unbounded ones $X^{n}$.
Notice the difference between the static LCR-structure (electron) ( \ref%
{g10a}) written below and the "static" LCR-structure in the bounded
realization%
\begin{equation}
\begin{array}{l}
(bounded)\qquad K_{B}^{(e)}=\frac{C}{2}(Y^{0}Y^{1}-Y^{2}Y^{3})+\frac{C-2D}{2}%
(Y^{1}Y^{2}-Y^{0}Y^{3}) \\ 
\multicolumn{1}{c}{\Updownarrow} \\ 
(unbounded)\qquad K_{U}^{(e)}=CX^{0}X^{1}+D(X^{1}X^{2}-X^{0}X^{3})%
\end{array}
\label{b7}
\end{equation}%
This difference of the asymptotic bounded and unbounded states may be the
origin of the neutrino (and quark) mixing, which appears in their time
evolution.

\bigskip

\end{document}